\documentclass{llncs}
\usepackage{cite}

\usepackage{graphicx}

\usepackage{comment}
\usepackage[inline]{enumitem}
\usepackage{listings}
\usepackage{color}

\usepackage{subfig}
\usepackage{array}
\usepackage{booktabs}
\usepackage{footnote}
\newcolumntype{P}[1]{>{\centering\arraybackslash}m{#1}}
\usepackage{enumitem}
\setitemize{noitemsep,topsep=0pt,parsep=0pt,partopsep=5pt}
\setenumerate{noitemsep,topsep=0pt,parsep=0pt,partopsep=2pt}
\setdescription{noitemsep,topsep=0pt,parsep=0pt,partopsep=4pt}
\makeatletter

\makeatletter
\def\@seccntformat#1{\@ifundefined{#1@cntformat}
   {\csname the#1\endcsname\quad}  
   {\csname #1@cntformat\endcsname}
}
\let\oldappendix\appendix 
\renewcommand\appendix{
    \oldappendix
    \newcommand{\section@cntformat}{\appendixname~\thesection\quad}
}
\makeatother

\begin{document}

\title{Querying Linked Data: An Experimental Evaluation of State-of-the-Art Interfaces}

\author{Gabriela Montoya \and Ilkcan Keles \and Katja Hose}

\institute{Aalborg University, Aalborg, Denmark\\ \email{\{gmontoya,ilkcan,khose\}@cs.aau.dk}}

\maketitle

\begin{abstract} 
	The adoption of Semantic Web technologies, and in particular the Open Data initiative, has contributed to the steady growth of the number of datasets and triples accessible on the Web.
Most commonly, queries over RDF data are evaluated over SPARQL endpoints. Recently, however, alternatives such as TPF have been proposed with the goal of shifting query processing load from the server running the SPARQL endpoint towards the client that issued the query. 
Although these interfaces have been evaluated against standard benchmarks and testbeds that showed their benefits over previous work in general, a fine-granular evaluation of what types of queries exploit the strengths of the different available interfaces has never been done. 
In this paper, we present the results of our in-depth evaluation of existing RDF interfaces. 
In addition, we also examine the influence of the backend 
on the performance of these interfaces. 
Using representative and diverse query loads based on the query log of a public SPARQL endpoint, we stress test the different interfaces and backends and identify their strengths and weaknesses. 
\end{abstract} 

\section{Introduction}
\label{sec:introduction}

With the adoption of the Open Data initiative by many institutions and companies, the amount of data offered on the Web in RDF is growing on a daily basis. While some of these datasets, such as DBpedia~\cite{DBLP:journals/semweb/LehmannIJJKMHMK15}, offer information extracted from unstructured sources, such as Wikipedia, other datasets focus on factual information from a specific domain, such as life science, geography, government, publications, etc. 

The simplest way to make such datasets available to others is publishing them on the Web as downloadable data dumps, typically encoding information in RDF formats such as N-triples or Turtle. The dump can then be downloaded through HTTP and the user can process the data according to his/her needs. Whereas this is very low effort for the data provider, the problem is that the user cannot simply query the information he/she is looking for directly at the data provider but instead has to download the entire dataset and process the query locally. 

On the other hand, a data provider can choose to run a SPARQL endpoint (server) to provide access to the data. In this way, a user (client) can send a SPARQL query to the endpoint, which processes it and returns the answer to the query. The advantage for the user in this setting is that it requires very little effort from the user. The price, however, is a relatively high load at the server running the endpoint as it has to process the entire query. If too many clients send queries concurrently or if the server is processing  complex queries, query response time increases and/or the endpoint might even become unavailable for some time.

To address this bottleneck, the Triple Pattern Fragments~\cite{DBLP:journals/ws/VerborghSHHVMHC16} (TPF) interface was proposed. To achieve the goal of better sharing the query load between clients and servers, the server is stripped from any higher-level query functionality and is only able to process single triple pattern requests. Any other query processing tasks, in particular processing joins, filters, grouping, are exclusively handled by the client. In doing so, the TPF interface increases availability and throughput at the server side. brTPF~\cite{DBLP:conf/otm/HartigA16} then extends TPF by allowing the client to not only send a single triple pattern to the server but also include a sequence of bindings for the variables in the triple pattern. This allows to include bindings obtained from intermediate results of a SPARQL query at the client and reduce the number of HTTP requests that need to be sent to the server.

Although the literature~\cite{DBLP:conf/otm/HartigA16,DBLP:conf/semweb/HartigLP17,DBLP:conf/semweb/MontoyaAH18,DBLP:journals/ws/VerborghSHHVMHC16} provides some analysis of the general behavior of available RDF interfaces, in particular SPARQL endpoints, TPF, and brTPF, none of them provides a detailed analysis that tries to find out which interface performs best for a specific query type and what the advantages of an interface over another interface are for processing a specific query type. Even though a formal framework for comparing RDF interfaces in terms of their expressiveness and complexity is proposed by Hartig et al.~\cite{DBLP:conf/semweb/HartigLP17}, there is no empirical evaluation addressing this question. 
In this paper, we therefore provide an extensive empirical evaluation of available RDF interfaces (SPARQL endpoint, TPF, and brTPF) using a real query load sent to the DBpedia SPARQL endpoint~\cite{soton385344}. In contrast to existing analyses, we decided to use a real query log instead of a synthetic benchmark to reflect real user behavior.

Another interesting aspect that is not covered by existing analyses is to what degree the overall performance of an RDF interface is actually determined by the chosen backend on the server. In other words, there is simply no study that analyzes the effect of the backend used on the performance of RDF interfaces. In the case of a SPARQL endpoint, the query is just processed by a triple store (Virtuoso, Jena, Blazegraph, etc.). TPF and brTPF, on the other hand, can operate with either a triple store or an HDT file~\cite{DBLP:journals/ws/FernandezMGPA13} as backend. 

In summary, this paper makes the following contributions: (i) A survey of coverage and shortcomings of existing evaluations of RDF interfaces, (ii) Definition of representative and diverse query loads that facilitates an in-depth analysis of RDF interfaces, and (iii) Extensive evaluation of RDF interfaces analyzing how much shapes of the queries influence the performance of available RDF interfaces (SPARQL endpoint, TPF, and brTPF) and what influence the backend has on the server's performance.

This paper is organized as follows.
Section~\ref{sec:existingEvaluations} summarizes existing evaluations of RDF interfaces and highlights their shortcomings. 
Section~\ref{sec:setup} defines the experimental setup, including the characterization of representative queries from logs of public endpoints. Section~\ref{sec:results} presents our experimental results and an extensive discussion, and 
finally Section~\ref{sec:conclusion} concludes the paper with a summarization of our most important findings.

\section{Existing Interfaces and Evaluations}
\label{sec:existingEvaluations}

In this paper, we focus on the most popular interfaces proposed for querying RDF datasets: SPARQL endpoints, Triple Pattern Fragments (TPF)~\cite{DBLP:journals/ws/VerborghSHHVMHC16}, and bindings-restricted Triple Pattern Fragments (brTPF)~\cite{DBLP:conf/otm/HartigA16}. 
SPARQL endpoints are most convenient for clients as they can submit complete SPARQL queries and simply receive the answer to the query -- the complete query load is on the server side (SPARQL endpoint). Furthermore, SPARQL endpoints support the complete SPARQL specification~\cite{clark13_sparql}.
On the other hand, allowing clients to issue complex SPARQL queries might require considerable resources in terms of CPU and main memory at the server. Processing multiple such queries concurrently might result in considerable delays or in the worst case non-availability of the server. 
A survey of public SPARQL endpoints~\cite{DBLP:conf/semweb/ArandaHUV13} shows that only $32.2$\% of endpoints are capable of providing $99$\% to $100$\% availability during the $27$-month long monitoring. 

The TPF interface~\cite{DBLP:journals/ws/VerborghSHHVMHC16} was proposed to address the availability issue of SPARQL endpoints by better sharing the query processing load between server and clients. 
A TPF server is only capable of handling triple pattern requests. In other words, it receives a triple pattern from a client and returns the triples of the hosted knowledge graph matching the input triple pattern. 
The client then takes care of all other query processing tasks, such as joining, filtering, grouping, query optimization and decomposition, and sending triple pattern requests to the servers.
The TPF interface has been evaluated against SPARQL endpoints based on Jena Fuseki and Virtuoso~\cite{DBLP:books/sp/virgilio09/ErlingM09} using an instance of the Berlin SPARQL Benchmark (BSBM) dataset~\cite{DBLP:journals/ijswis/BizerS09} that contains 100 millions triples. 
The experiments show that the CPU load on the server is lower and the CPU load on the client is higher for TPF interfaces compared to SPARQL endpoints. Moreover, the network load between the server and the client increases since the client has to issue several HTTP requests to process a single SPARQL query. Verborgh et al.~\cite{DBLP:journals/ws/VerborghSHHVMHC16} also provide an experiment to assess the performance of TPF on a real-world knowledge graph by executing different queries obtained from the DBpedia SPARQL benchmark (DBSB)~\cite{DBLP:conf/semweb/MorseyLAN11} on three knowledge bases containing 14 million triples, 52 million triples, and 377 million triples, respectively. This experiment shows that the query processing time of TPF has a high variance between queries with different keywords. Queries with keywords like UNION, FILTER, and OPTIONAL are quite expensive using a TPF client since a TPF client does not provide a good query plan for such queries. TPF was the only interface assessed in this experiment and no results regarding the execution of these queries against SPARQL endpoints were provided. Moreover, the experiments presented do not pay any particular attention to the type of the issued query. For this reason, the effect of the issued query type on TPF remains unknown.

brTPF~\cite{DBLP:conf/otm/HartigA16} extends TPF by adding a sequence of bindings to the triple pattern requests to reduce the overall number of HTTP requests necessary to answer a query. brTPF was evaluated against TPF using the WatDiv dataset and queries generated by the associated stress testing tools~\cite{DBLP:conf/semweb/AlucHOD14}. Specifically, a synthetic knowledge graph with 10 million triples (published also on the project website) are used for evaluation. A total of $145$ BGP queries and a total of $12,400$ queries are used for single-client experiments and multi-client experiments, respectively. Up to $64$ clients are used for multi-client experiments. The experimental evaluation demonstrates that brTPF has a better query throughput and less network overhead compared to TPF in both settings. 

Besides from proposing WatDiv dataset and stress testing tools, Aluc et al.~\cite{DBLP:conf/semweb/AlucHOD14} also present an experimental evaluation of SPARQL endpoints including Virtuoso and 4store. The experimental evaluation shows that the query processing performance of the endpoints differs a lot with respect to the queries. In order to see the effect of query characteristics on the performance of the endpoints, they group queries with respect to their selectivity and their structure. They only consider linear and star/snowflake structures. According to the reported evaluation results, no single SPARQL endpoint is the best or the worst across all queries. For this reason, we believe that it is also important to systematically analyze how the type of the issued query affects the RDF interfaces. 

Both lines of experiments ~\cite{DBLP:conf/otm/HartigA16,DBLP:journals/ws/VerborghSHHVMHC16} are based on servers using HDT files~\cite{DBLP:journals/ws/FernandezMGPA13} as backend. HDT (Header, Dictionary, Triples) is proposed as a binary serialization format for RDF data. It is shown to compress the data significantly while still allowing for efficient query processing~\cite{DBLP:journals/ws/FernandezMGPA13}.
Therefore, the effect of the backend triple store, in particular a standard triple store as an alternative to HDT, on the performance of TPF and brTPF remains unknown. In other words, whether TPF and brTPF provide any improvement when using standard triple stores as backend has so far not yet been analyzed. We think that this is quite important since the main target of TPF and brTPF is reducing the load on the servers and as a consequence increasing the availability of the triple stores hosting the RDF datasets. 
Moreover, existing evaluations between TPF and brTPF are limited to WatDiv and do not analyze the influence of particular types of queries. Instead, only average times over sets of queries are reported. 
However, a solution that works well on average does not necessarily work best on all types of queries.

\section{Evaluation Setup}
\label{sec:setup}

In this section, we present our experimental setup covering datasets and queries, query loads, interfaces, hardware setup, and evaluation metrics.

\textbf{Dataset and Queries. }
For our study, we use the DBpedia 3.9 dataset\footnote{Available at \url{http://downloads.dbpedia.org/3.9/en/}} and the USEWOD 2016 research dataset~\cite{soton385344} that contains query logs from the public DBpedia interfaces: SPARQL endpoint and TPF server. We use SPARQL queries sent to the DBpedia SPARQL endpoint. The USEWOD dataset covers the query logs of $20$ randomly selected days between between July, 2015 and November, 2015 ($43$ GBs) containing nearly $10$ million unique select queries. We do not use existing benchmarks that generate synthetic queries such as~\cite{DBLP:conf/semweb/AlucHOD14,DBLP:journals/ijswis/BizerS09} since the generated queries do not sufficiently reflect the characteristics of queries executed by actual users of the SPARQL endpoints. Moreover, we also do not use existing benchmarks based on user query logs such as~\cite{DBLP:conf/semweb/MorseyLAN11,DBLP:conf/semweb/SaleemMN15} since they focus on a set of queries covering all possible SPARQL keywords while we want to focus on BGPs, OPTIONALs and FILTERs in order to perform a fair comparison between the interfaces as TPF and BrTPF do not support some keywords.

A recent study~\cite{DBLP:journals/pvldb/BonifatiMT17} provides structural and shape analysis for all the queries in the USEWOD 2016 research dataset~\cite{soton385344}. According to this analysis, six shapes are the most common shapes in query logs: CHAIN, CYCLE, EDGE, FLOWER, TREE, and STAR. For our study, we only consider unique select queries that do not have any syntactical errors according to the SPARQL specification and whose shape corresponds to one of the above mentioned shapes. Moreover, we also consider structural characteristics such as the use of operators JOIN, OPTIONAL, and FILTER, use of variables as predicates, whether the used filters are safe and simple and whether the used OPTIONAL clauses are well-designed and tractable in line with Bonifati et al.~\cite{DBLP:journals/pvldb/BonifatiMT17}. A safe filter only includes variables used in its graph pattern, while simple filters include only one variable or correspond to $X=Y$ with $X$, $Y$ being variables. Well designed OPTIONAL clauses only join graph patterns using variables that are always bound (in the left operand), while tractable optionals include at most one join variable between their operands. In this study, we focus on the queries that do not have any variables as predicates, and that contains only safe simple FILTER clauses and well-designed and tractable OPTIONAL clauses. In other words, we focus on tractable queries and there are $4,337,181$ such queries contained in the USEWOD 2016 dataset. 

\begin{table}[htb]
\vspace*{-3.25em}
\centering
\caption{Shapes of Queries}
\begin{tabular}{c|r|r}
\toprule
\textbf{Query Shape} & \textbf{Total} & \textbf{Relevant}\\ \hline
 CHAIN & 832,873 & 171,244\\\hline
 CYCLE & 73 &  31\\\hline
 EDGE & 3,189,874 & 1,275,313 \\\hline
 FLOWER & 3,209 & 5 \\\hline
 STAR & 274,678 & 8,657 \\\hline
 TREE & 36,474 &  358\\\hline
\end{tabular}
\label{tab:shapes}
\vspace*{-2em}
\end{table}

For these 4,337,181 queries, we examined their shapes and the findings are listed in Table~\ref{tab:shapes}. The total number corresponds to the number of queries with this shape. We exclude queries with empty answers and queries that are not supported by existing implementations to allow for a more interesting performance study of existing RDF interface implementations using these queries. Existing implementations of TPF and brTPF do not support features such as VALUES, subqueries, REGEX expressions with three arguments, aggregations, functions on RDF terms (e.g., isLiteral), and functions on strings (e.g., UCASE). Moreover, some predicates such as \textit{bif:contains} are only supported by Virtuoso. We refer to the remaining queries as \textit{relevant queries}.
The number of relevant queries for our study is shown in the rightmost column of Table~\ref{tab:shapes}. 
Some query shapes had considerably fewer queries that have answers and are supported by existing implementations. An example is queries with FLOWER shape, where 3,108 out of 3,209 queries used the predicate bif:contains that is only supported by Virtuoso. After we determine the set of relevant queries, we remove modifiers DISTINCT, ORDER BY, LIMIT and OFFSET from the queries with these modifiers. This is needed to focus on the evaluation of the graph patterns and to have a fair comparison between different interfaces. Since both TPF and brTPF are not optimized for modifiers and use post-processing on the client-side to process queries with modifiers, we think it would be unfair to compare brTPF and TPF with SPARQL endpoints using such queries. 

\textbf{Query Loads. }
After determining the set of relevant queries, we continue with creating query loads for single-client and multiple-clients experiments. In line with~\cite{DBLP:conf/otm/HartigA16}, we include experimental evaluation with multiple clients to assess how the number of clients concurrently accessing the interface effects the performance of the interface. Moreover, this set of experiments makes it possible to evaluate the interfaces under high load.

For the single-client experiments, we consider query loads of $100$ random queries for each shape\footnote{Except for shapes with less than $100$ queries. In this case, all the available queries are included in the query load.}. In total, we have $443$ queries distributed into $6$ query loads, $1$ for each shape: CHAIN, CYCLE, EDGE, FLOWER, TREE, and STAR. 

For multiple-client experiments, instead of creating a separate query load for each shape, we create two query loads (Equal and Proportional) for each client that combine queries with different shapes. Both query loads are constructed by randomly drawing queries from different query shapes. The queries are drawn with respect to the uniform distribution for the Equal query load and with respect to the frequency distribution for the Proportional query load. In the uniform distribution, every shape has the same probability to be drawn, while in the frequency distribution the shape probability is proportional to the frequency of that shape in the relevant queries. 

We want to make sure that the intersection of the query loads for different number of clients is empty since we do not want interfaces take advantage of caching during the experiments. For this reason, we considered only shapes with at least $6,400$ queries as we aimed to have $64$ query loads with $100$ queries to have experiments with $64$ clients.

\textbf{Interfaces and Backends. }
In this paper, we focus on the three standard interfaces for accessing RDF datasets: TPF~\cite{DBLP:journals/ws/VerborghSHHVMHC16}, brTPF~\cite{DBLP:conf/otm/HartigA16}, and SPARQL endpoints. 
We evaluate each of them in combination with $4$ different backends: HDT~\cite{DBLP:journals/ws/FernandezMGPA13}, Fuseki\footnote{part of hdt-java, available at \url{https://github.com/rdfhdt/hdt-java} latest development version, February 17th, 2019}, Virtuoso~\cite{DBLP:books/sp/virgilio09/ErlingM09} 7.2.5.3229-pthreads, and Blazegraph~\cite{DBLP:books/crc/linked14/ThompsonPC14} 2.1.5 Release Candidate version. Fuseki is used with default configuration\footnote{The configuration files for Virtuoso and Blazegraph are available at \url{http://qweb.cs.aau.dk/evaluation}.}.

We use the nodeJS client from~\cite{DBLP:conf/otm/HartigA16} for brTPF, the nodeJS client from~\cite{DBLP:journals/ws/VerborghSHHVMHC16} for TPF, and a nodeJS client from ~\cite{DBLP:conf/semweb/MontoyaAH18} for SPARQL endpoints. 
There are different clients that might be used for querying against SPARQL endpoints. However, we use a nodeJS client since we want to have a fair comparison between interfaces by relying on the same client infrastructure. Thus, we aim to reduce any difference that could be attributed for instance to programming languages or different result formats.

We extended the Java TPF server\footnote{\url{https://github.com/LinkedDataFragments/Server.java}} with support for brTPF requests and additional SPARQL-based backends such as Virtuoso endpoints\footnote{The code is available at \url{http://qweb.cs.aau.dk/evaluation/brTPF-server-ISWC.zip}.} and use that as the server component for TPF and brTPF.

\begin{table*}[htb]
\vspace*{-3.5em}
\centering
\caption{Machine configurations}
\begin{tabular}{c|c|c|c}
\toprule
\textbf{Machine} & \textbf{Cores} & \textbf{RAM} & \textbf{OS}\\ \hline
Small & 8 x 2294.250 MHz &  64GB & Ubuntu 16.04.1 LTS \\\hline
Big   & 64 x 2294.176 MHz & 516GB & Ubuntu 14.04.6 LTS \\\hline
\end{tabular}
\label{tab:machines}
\vspace*{-2em}
\end{table*}

\textbf{Hardware Configuration. }
We use two machines with different configurations that are described in Table~\ref{tab:machines}. 
For the experiments with a single client, the servers are deployed using docker\footnote{https://www.docker.com/} containers in the big server and configured so that they will use up to 8GB of RAM and three cores, while the clients are run in the small machine and each client is set to use up to 3GB of RAM. 
For the experiments with multiple clients, the servers are deployed using docker containers in the small machine and configured to use up to 21GB of RAM\footnote{Virtuoso was not able to handle 64 clients with less RAM and lower bounds set by the configuration file failed to have any impact on restricting the RAM usage.} and three cores, while up to 64 clients are run in the big machine and each client is set to use up to 3GB of RAM.

\textbf{Evaluation Metrics. }
In the experimental evaluation we refer to SPARQL endpoints, TPF server and brTPF server as \textit{server}, and we refer to backends for TPF and brTPF as \textit{backend}. To evaluate the different approaches, we use the following measures:
\begin{itemize}
  \item Execution Time (ET): the time elapsed between issuing the query and getting the query results (with a timeout of five minutes),
  \item Number of HTTP requests (NH): the number of HTTP requests sent to the the server,
  \item Server Load (SL): the CPU percentage used by the TPF server, brTPF server, and SPARQL endpoints during query processing. It might go up to 300\% (as each server has 3 cores),
  \item Backend Load (BL): the CPU percentage used by the backend for TPF and brTPF during query processing. It might go up to 300\% (as each backend has 3 cores),
  \item Number of Retrieved kBs (NRKB): the number of kilobytes transferred from the server to the client during query execution,
  \item Number of Sent kBs (NSKB): the number of kilobytes transferred from the client to the server during query execution,
  \item Number of Timed out Queries (NTQ): the number of queries that do not complete their execution within five minutes.
\end{itemize}

\section{Evaluation Results}
\label{sec:results}

In this section, we present the results of the single-client and multiple-client experiments. We performed experiments using $11$ different combinations  of interfaces and backends in total: 
\begin{itemize}
	\item Endpoint interface: Blazegraph endpoint (e\textunderscore B), Fuseki endpoint (e\textunderscore F), and Virtuoso endpoint (e\textunderscore V), 
	\item brTPF interface: brTPF server with Blazegraph backend (b\textunderscore B), brTPF server with Fuseki backend (b\textunderscore F), brTPF server with HDT file backend (b\textunderscore H), and brTPF server with Virtuoso backend (b\textunderscore V),
	\item TPF interface: TPF server with Blazegraph backend (t\textunderscore B), TPF server with Fuseki backend (t\textunderscore F), TPF server with HDT file backend (t\textunderscore H), and TPF server with Virtuoso backend (t\textunderscore V).
\end{itemize}

\subsection{Preliminary Experiments}
\begin{figure*}[htb]
  \vspace*{-2em}
  \centering
  \subfloat[Errors]{
    \includegraphics[width=0.32\textwidth]{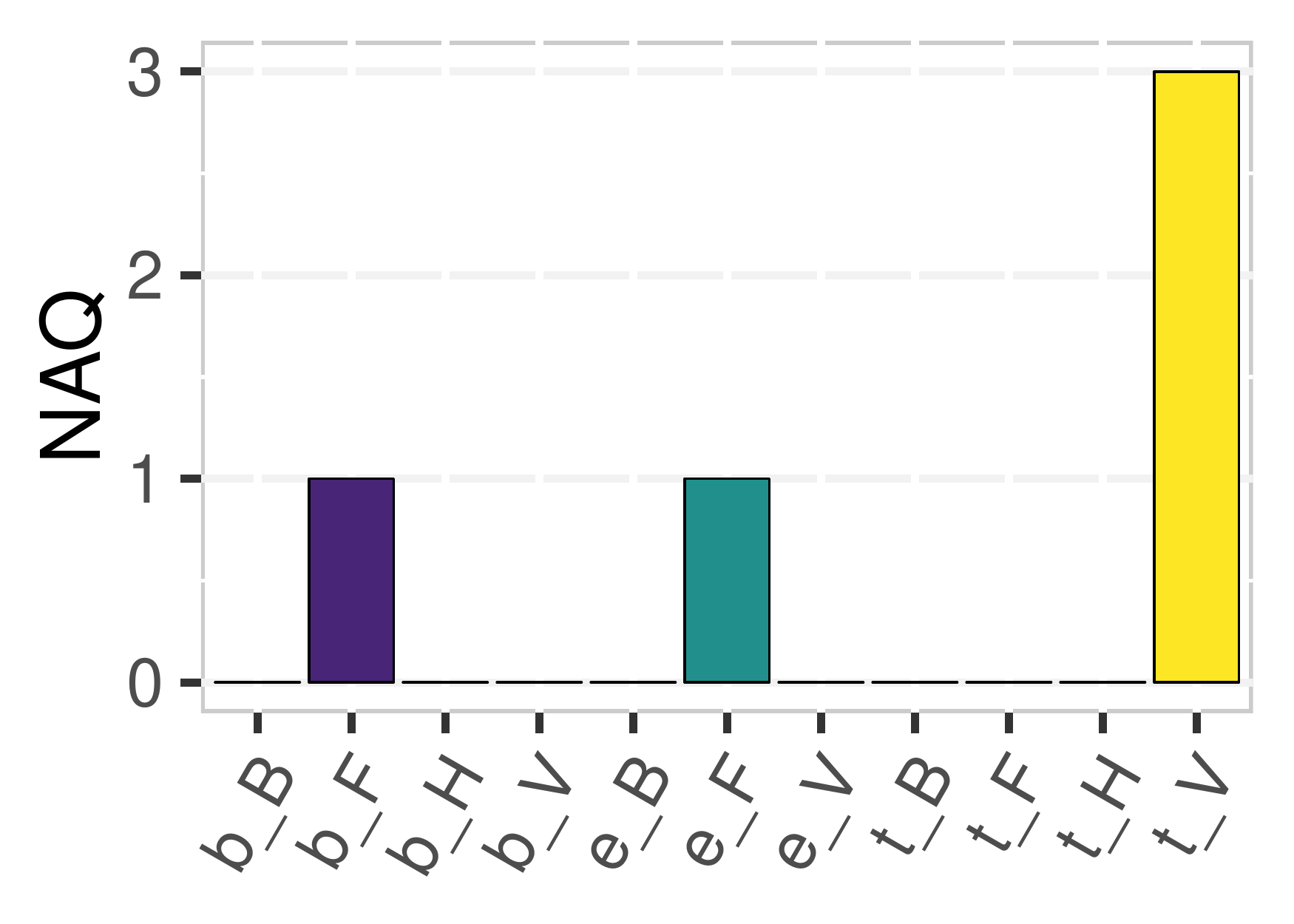}
  }
  \subfloat[Timeouts]{
    \includegraphics[width=0.32\textwidth]{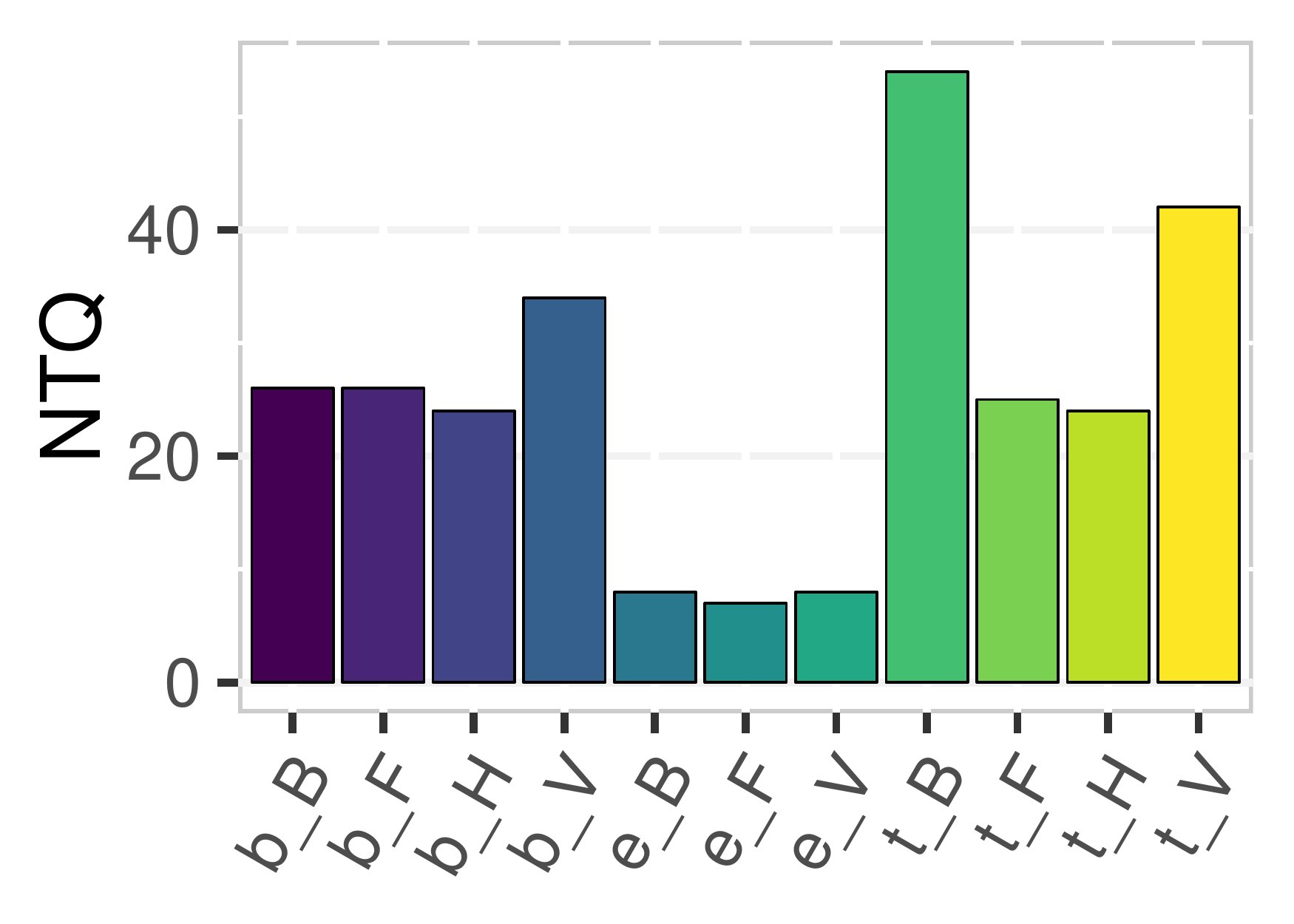}
  }
  \subfloat[Answers]{
    \includegraphics[width=0.32\textwidth]{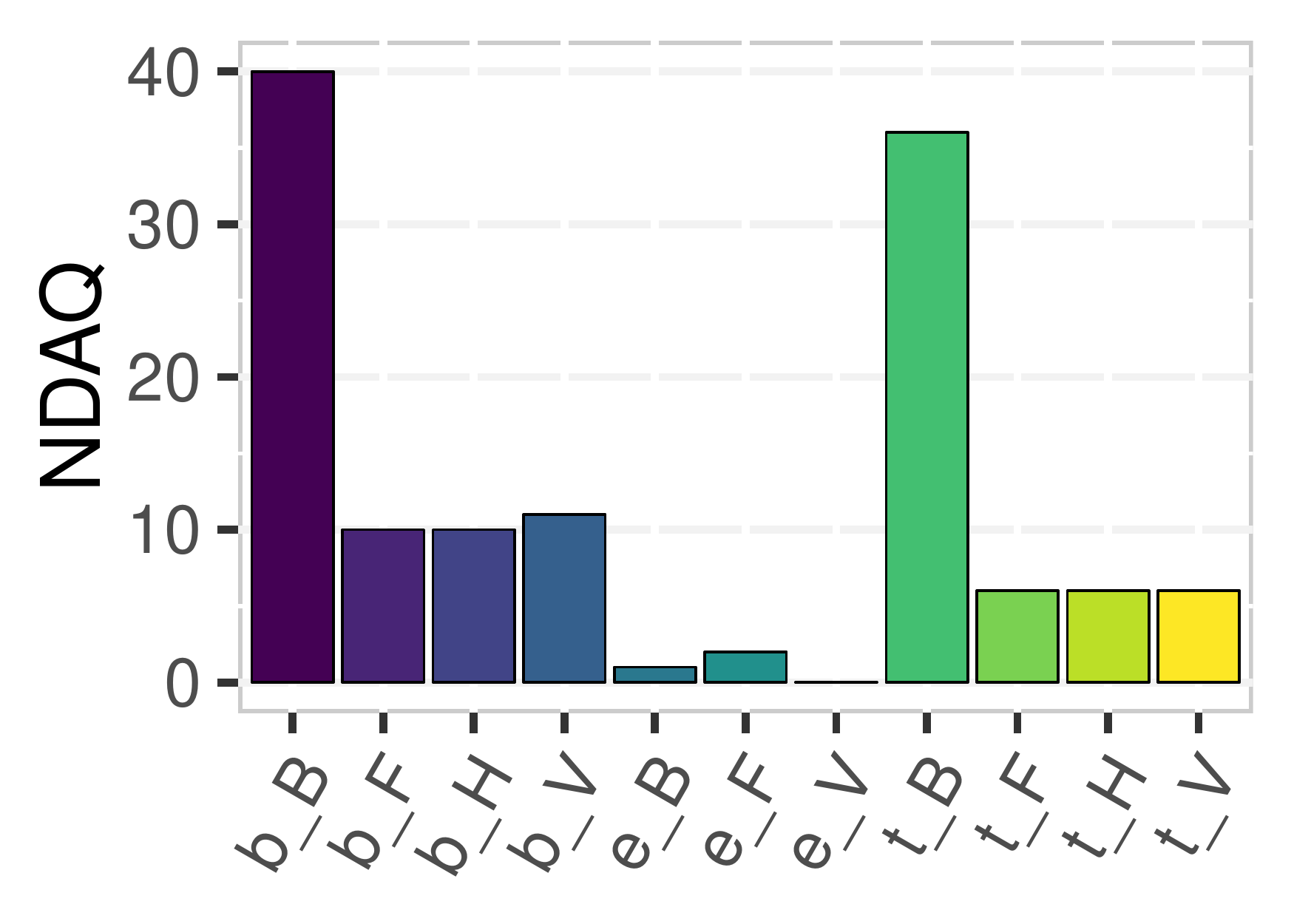}
  }
    \caption{Number of queries aborted with error, timed out, or with different number of answers}
\label{fig:preprocessing}
\end{figure*}

Surprisingly, we encountered several problems in our preliminary experiments when executing the generated query loads: queries that aborted with errors, queries with inconsistent results across systems, and timed-out queries. Figure~\ref{fig:preprocessing} shows an overview of such queries.
Including aborted and timed-out queries in our results can negatively impact the performance, data transfer, and server usage metrics of the systems that completed the execution of the queries without any problems (Appendix~\ref{sec:appA} shows the effect of including queries with consistent answers). Hence, we present the metrics obtained by excluding the problematic queries.

We assessed the reasons why we obtain different results across systems. 
The most common issue with TPF and brTPF is that their current implementations assume that the backend always provides the same result for the same pair of LIMIT and OFFSET values. However, 
Blazegraph can yield inconsistent results in different executions of the same query with the same pair of LIMIT and OFFSET values\footnote{\url{https://github.com/LinkedDataFragments/Server.js/issues/24}}. 
Other problems include incorrect evaluation of queries with OPTIONALs involving BGPs with more than one triple pattern (TPF), 
incorrect evaluation of queries with nested OPTIONALs (TPF and brTPF), fragment pages missing control elements that prevent accessing the second page of a fragment (brTPF), evaluation of property paths that do not follow the standard set semantics (Virtuoso). 

It is important to note that queries with the lowest number of triple patterns, e.g., EDGE-shaped queries, have no queries with different answers across the systems, while queries with higher numbers of triple patterns, e.g., TREE-shaped or STAR-shaped queries, amount to 41 of the total 45 queries (across all query loads) with different answers across systems. Therefore, studying queries with a higher numbers of triple patterns and diverse shapes allows for identifying some limitations of existing implementations of the different interfaces. 

\begin{figure*}[htb]
  \vspace*{-2em}
  \centering
  \subfloat[TPs]{
    \includegraphics[width=0.32\textwidth]{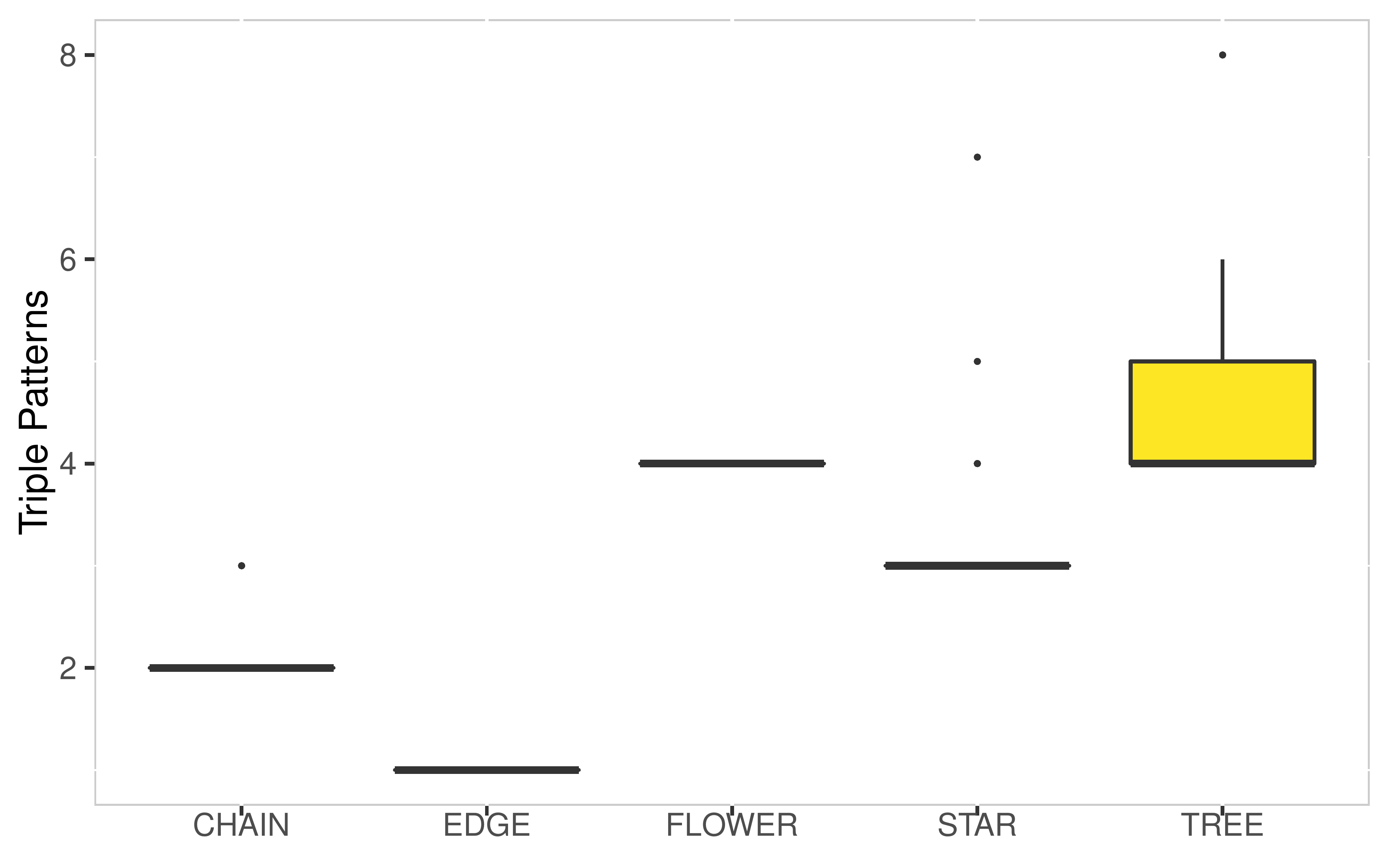}
  }
  \subfloat[BGPs]{
    \includegraphics[width=0.32\textwidth]{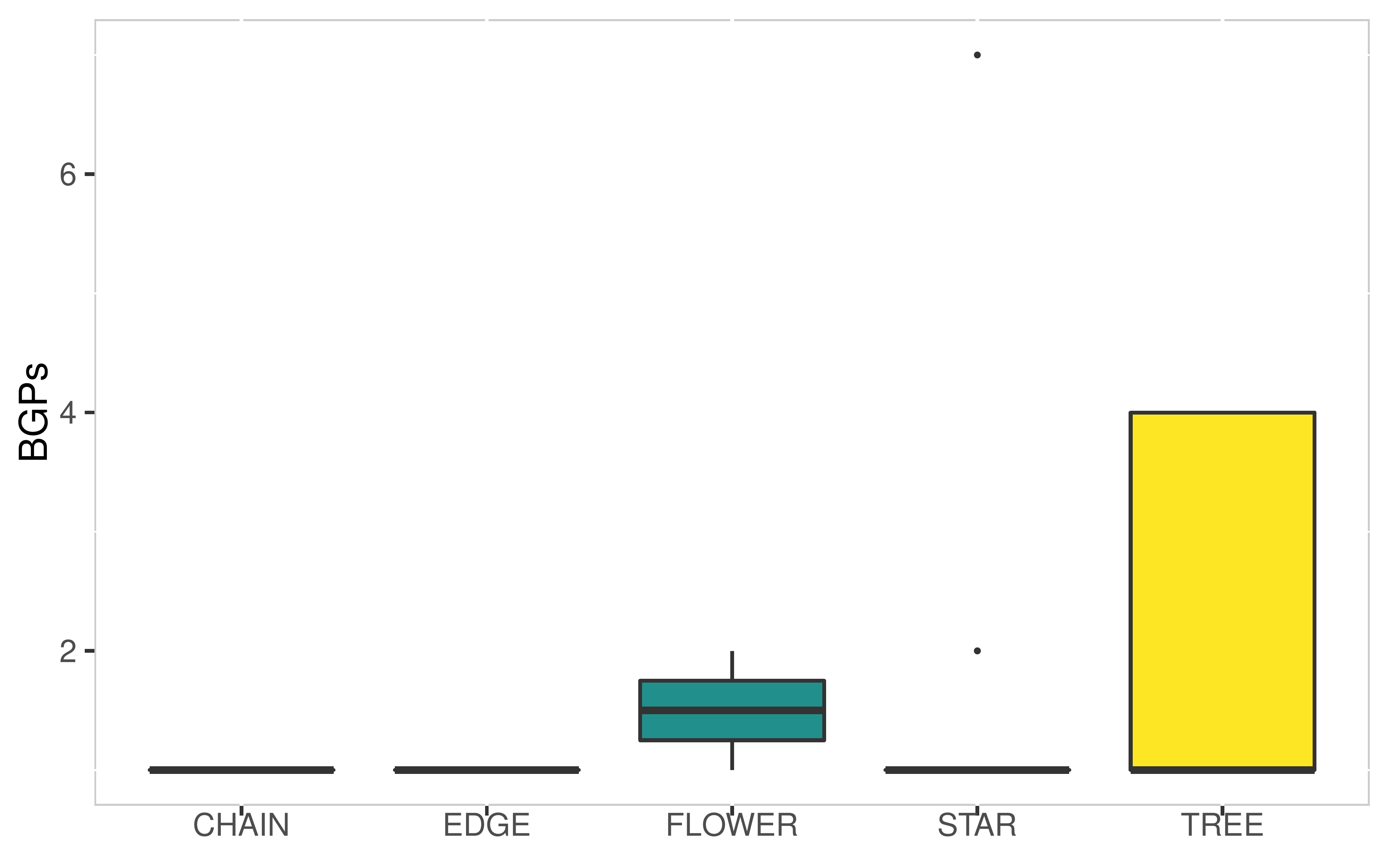}
  }
  \subfloat[Optionals]{
    \includegraphics[width=0.32\textwidth]{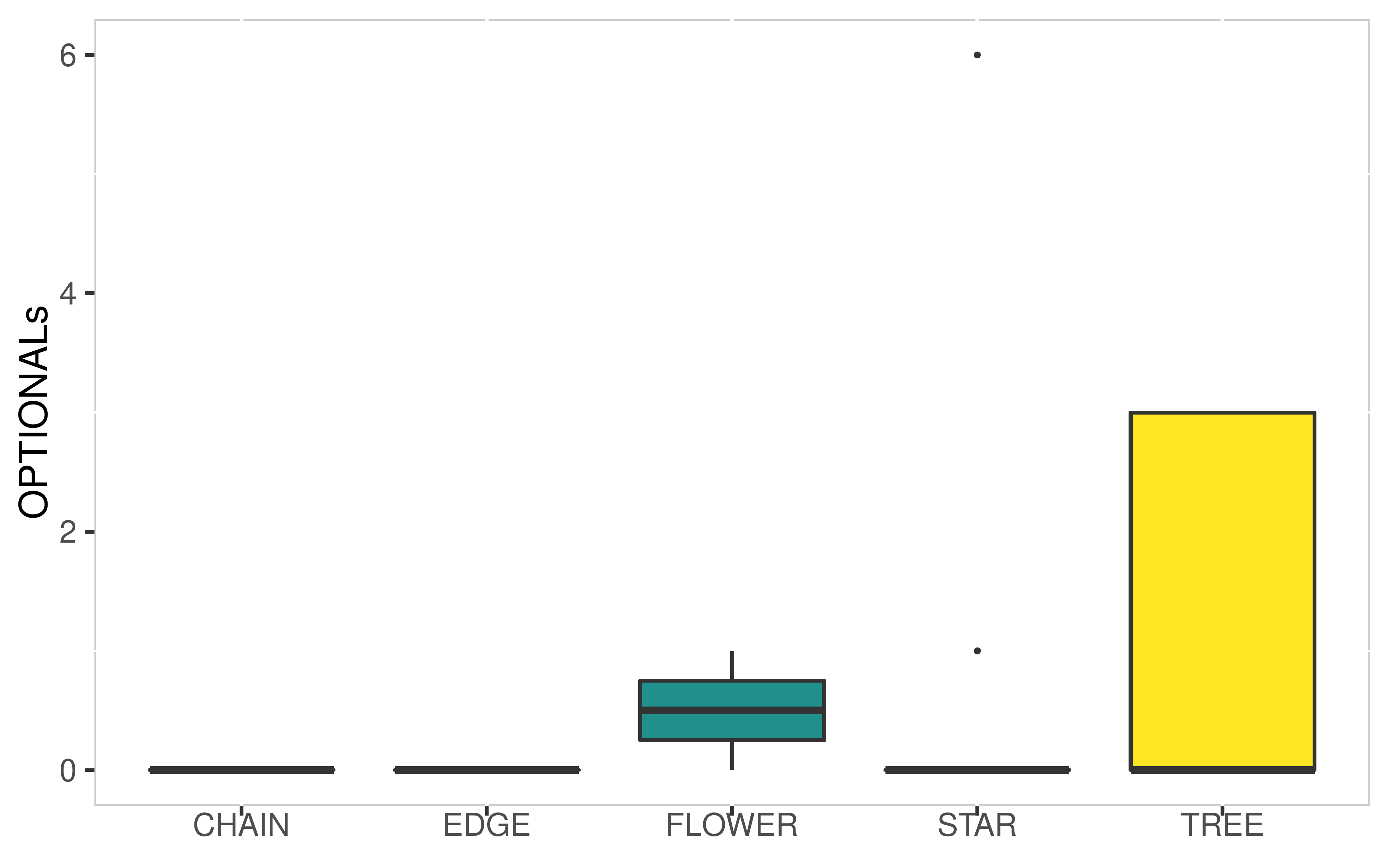}
  }\\
  \subfloat[Filters]{
    \includegraphics[width=0.32\textwidth]{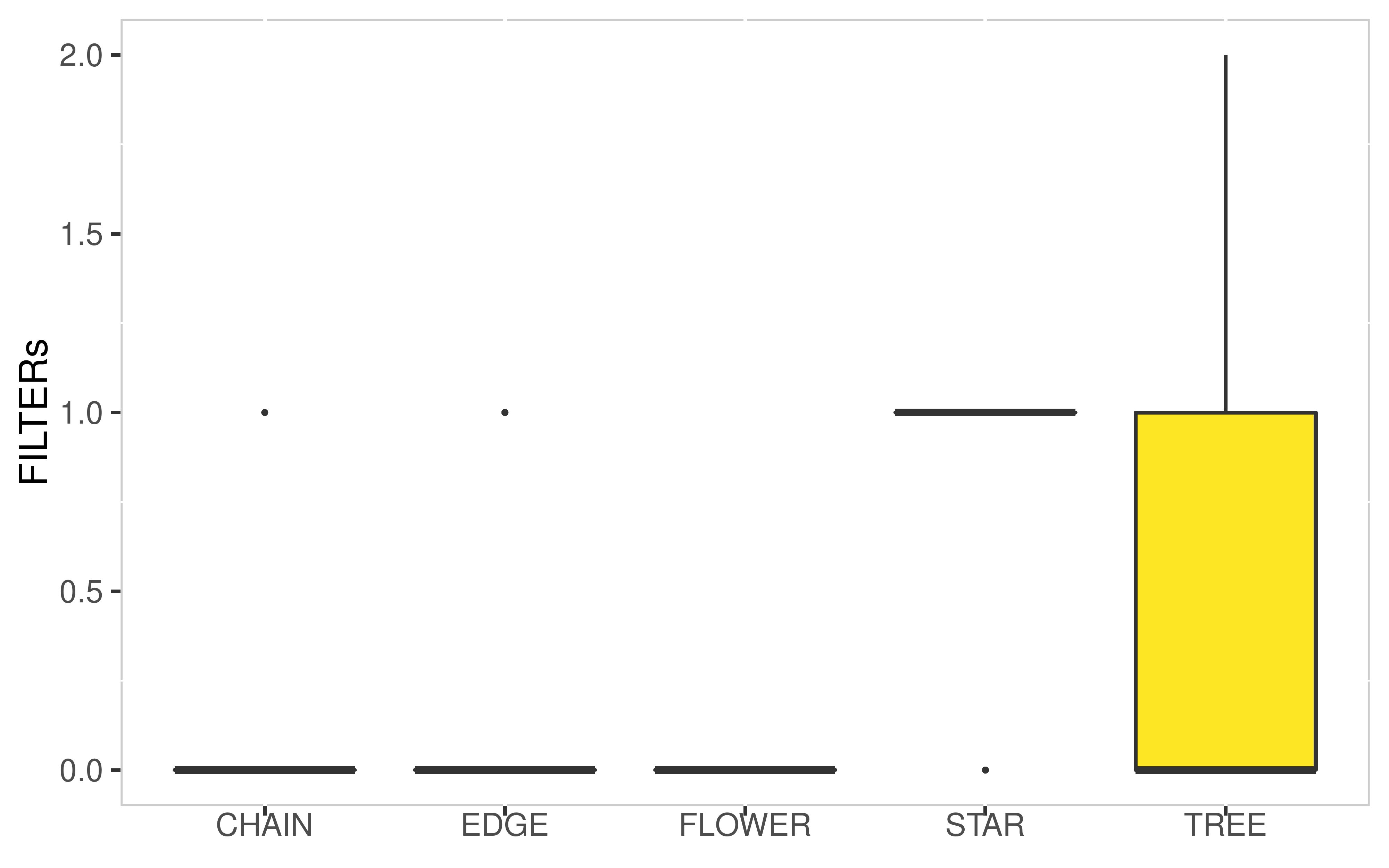}
  }
  \subfloat[Mean BGP Sel]{
    \includegraphics[width=0.32\textwidth]{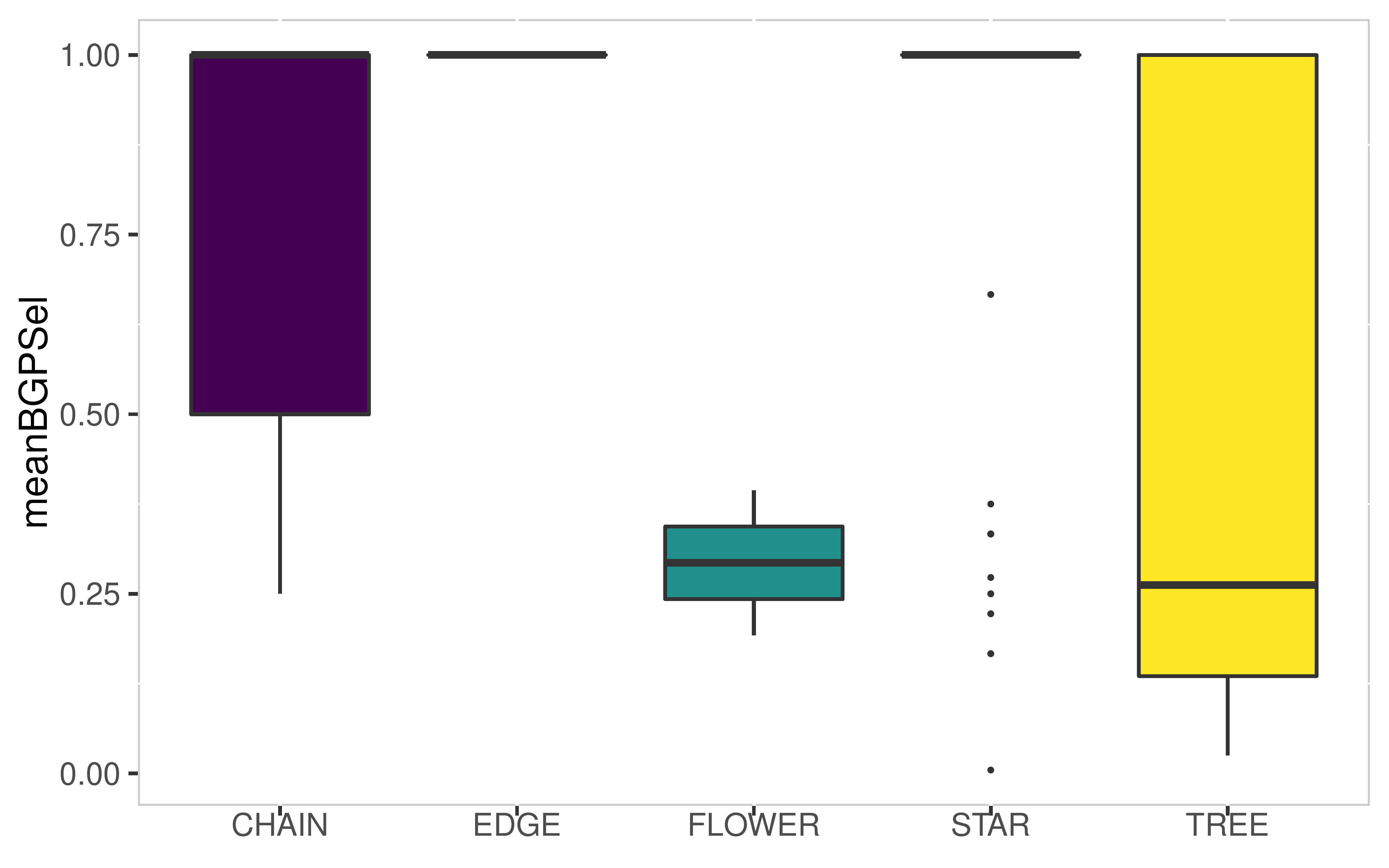}
  }
  \subfloat[Stdev BGP Sel]{
    \includegraphics[width=0.32\textwidth]{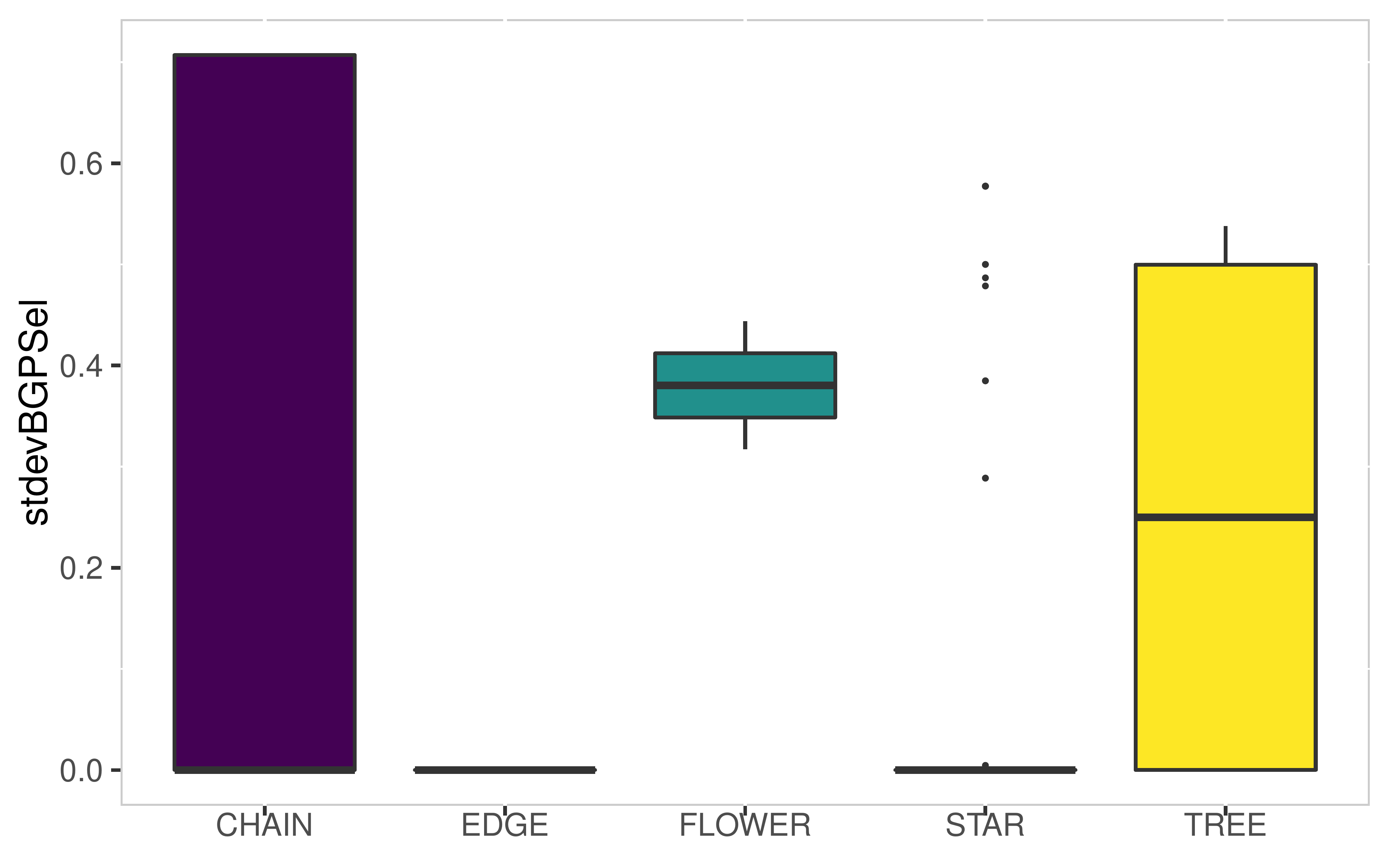}
  }
    \caption{Number of Triple Patterns (TPs), Basic Graph Patterns (BGPs), Optionals, and Filters, Mean and Standard Deviation (Stdev) BGP Selectivity (Sel)}
    \vspace*{-1.75em}
\label{fig:loadCharacteristics}
\end{figure*}

After removing the $45$ queries mentioned above, Figure~\ref{fig:loadCharacteristics} shows some structural and data-driven characteristics of the queries in each query load. 
The query load with higher diversity for these characteristics is TREE. It includes the higher number of OPTIONAL clauses and consequently the higher number of BGPs. 
It also includes queries with more diverse BGP selectivity. For a BGP $\mathit{bgp}=\lbrace \mathit{tp_1}, \mathit{tp_2}, ... , \mathit{tp_n} \rbrace$, the BGP selectivity of $\mathit{bgp}$ for $\mathit{tp_i}$ indicates the proportion of solutions for $\mathit{tp_i}$ that are compatible with solutions for $\mathit{bgp}$. 
A high BGP selectivity value indicates that most intermediate results contribute to the solution of $\mathit{bgp}$, while a low value indicates that there are many intermediate results that do not contribute to the solution of $\mathit{bgp}$. 

\subsection{Single-Client Experiments}
\begin{figure*}[htb]
\vspace*{-2em}
  \centering
  \subfloat[ET]{
    \includegraphics[width=0.32\textwidth]{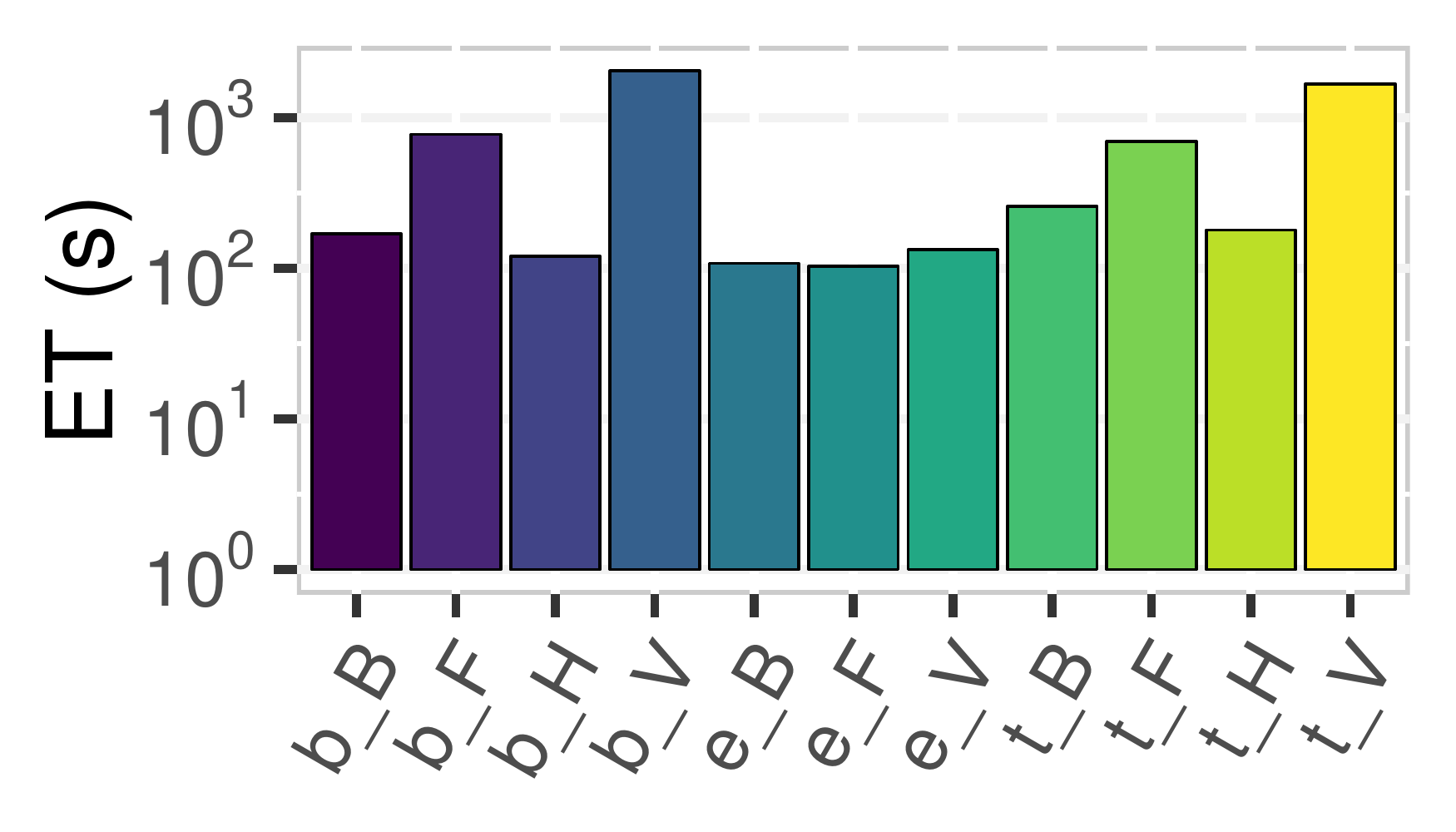}
  }
  \subfloat[QpS]{
    \includegraphics[width=0.32\textwidth]{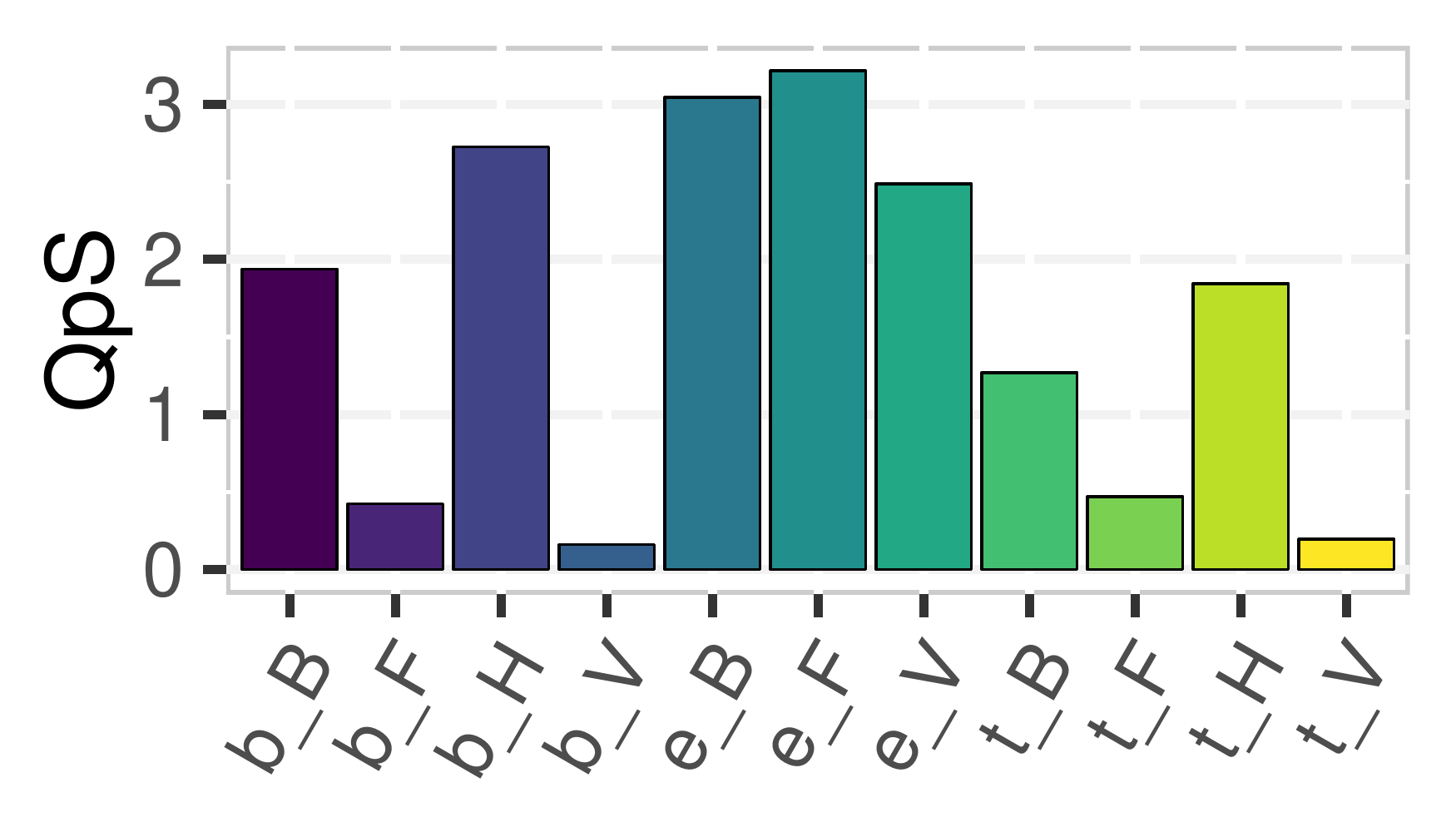}
  }
  \subfloat[QMpH]{
    \includegraphics[width=0.32\textwidth]{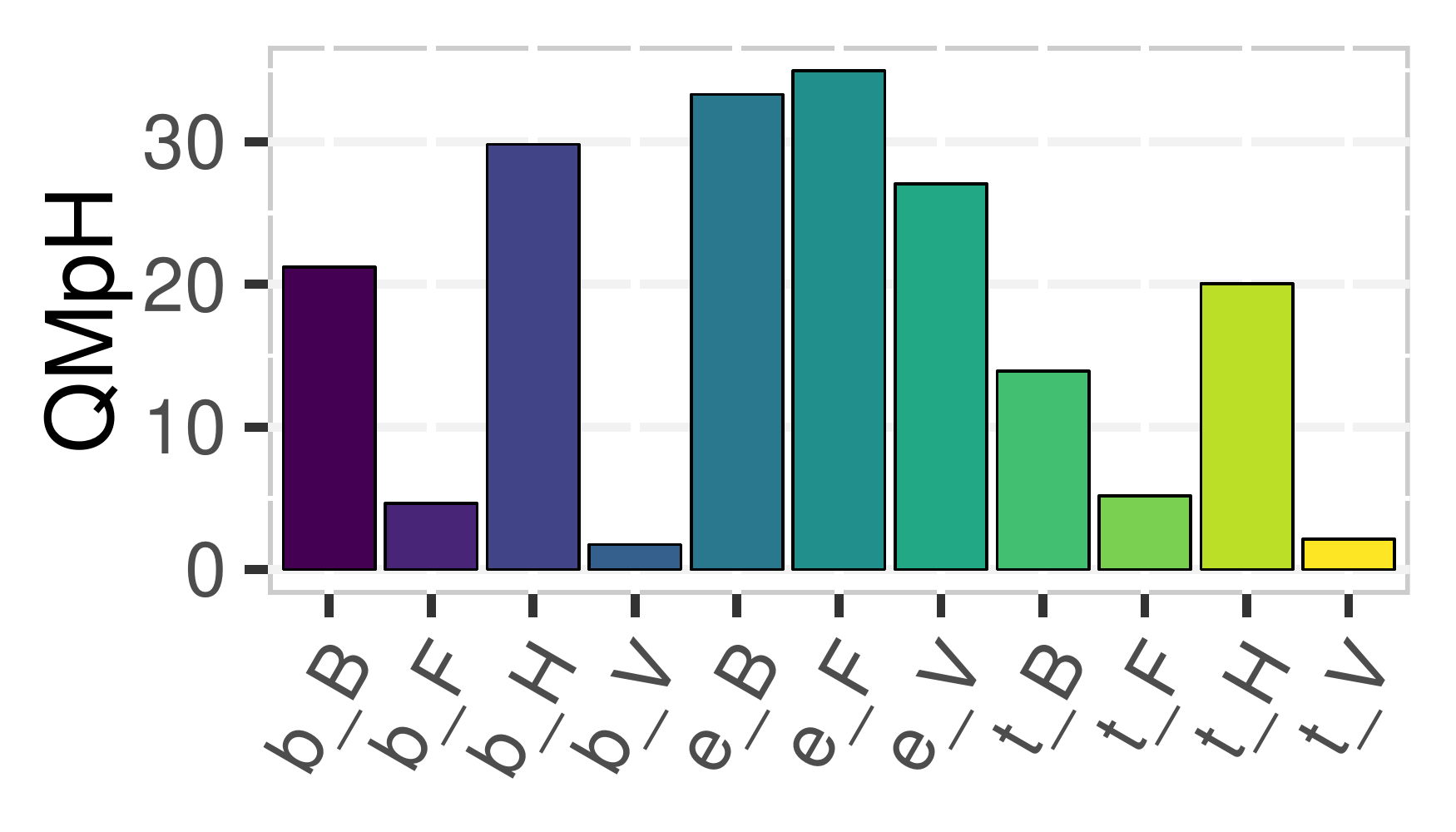}
  }
    \caption{Total ET, Queries per Second (QpS), and Query Mixes per Hour (QMpH) for different interfaces}
    \vspace*{-2em}
\label{fig:benchmarkMetrics}
\end{figure*}

\textbf{Performance. }
Existing benchmarks for SPARQL endpoints~\cite{DBLP:journals/ijswis/BizerS09,DBLP:conf/semweb/MorseyLAN11,DBLP:conf/semweb/SaleemMN15} use metrics such as queries per second (QpS) and query mixes per hours (QMpH) for performance evaluation. Existing TPF and brTPF studies~\cite{DBLP:conf/otm/HartigA16,DBLP:journals/ws/VerborghSHHVMHC16} employ queries per hour or throughput (QpH) metrics. 
All these metrics provide information with a very coarse granularity, i.e., just one number to describe how a system performed a query load. 
Figure~\ref{fig:benchmarkMetrics} shows some of these metrics for processing all single-client query loads. 
According to these results, the systems that perform the best are the Blazegraph and Fuseki endpoints (e\textunderscore B and e\textunderscore F). Moreover, the best performing backends for brTPF and TPF appear to be Blazegraph and HDT file while the worst one appears to be Virtuoso.

\begin{figure}[htb]
\vspace*{-1.5em}
  \centering
    \includegraphics[width=\textwidth]{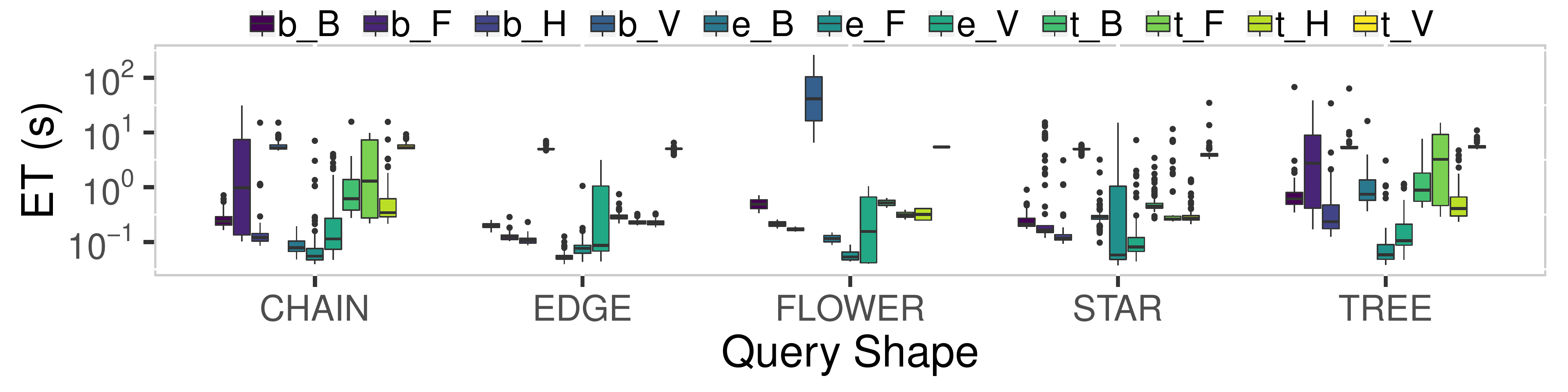}
    \vspace*{-1em}
    \caption{ET per query shape for interfaces}
\label{fig:detailedPerformanceSingleClient}
\vspace*{-2em}
\end{figure}

However, having a single number that summarizes the performance of the systems across the query loads may hide some interesting facts. 
Figure~\ref{fig:detailedPerformanceSingleClient} therefore shows the query execution time (ET) represented with a boxplot for each query shape and system. For instance, for queries with STAR shape, brTPF with any backend except by Virtuoso performs considerably better than the Blazegraph endpoint. Additionally, brTPF and TPF with HDT backend perform considerably better than Blazegraph endpoint for TREE-shaped queries. Moreover, the Virtuoso endpoint performs better than the Blazegraph endpoint for query shapes STAR and TREE. If we want to execute queries with characteristics as diverse as the ones in query load TREE (see Figure~\ref{fig:loadCharacteristics}), one would not choose
the Blazegraph endpoint even if it has the best overall performance according to Figure~\ref{fig:benchmarkMetrics}. Figure~\ref{fig:detailedPerformanceSingleClient} also shows that the shape of the issued queries affects the query processing performance on the interface / backend combinations. 

\begin{figure*}[!h]
\vspace*{-2em}
  \centering
  \subfloat[NH]{
    \includegraphics[width=0.32\textwidth]{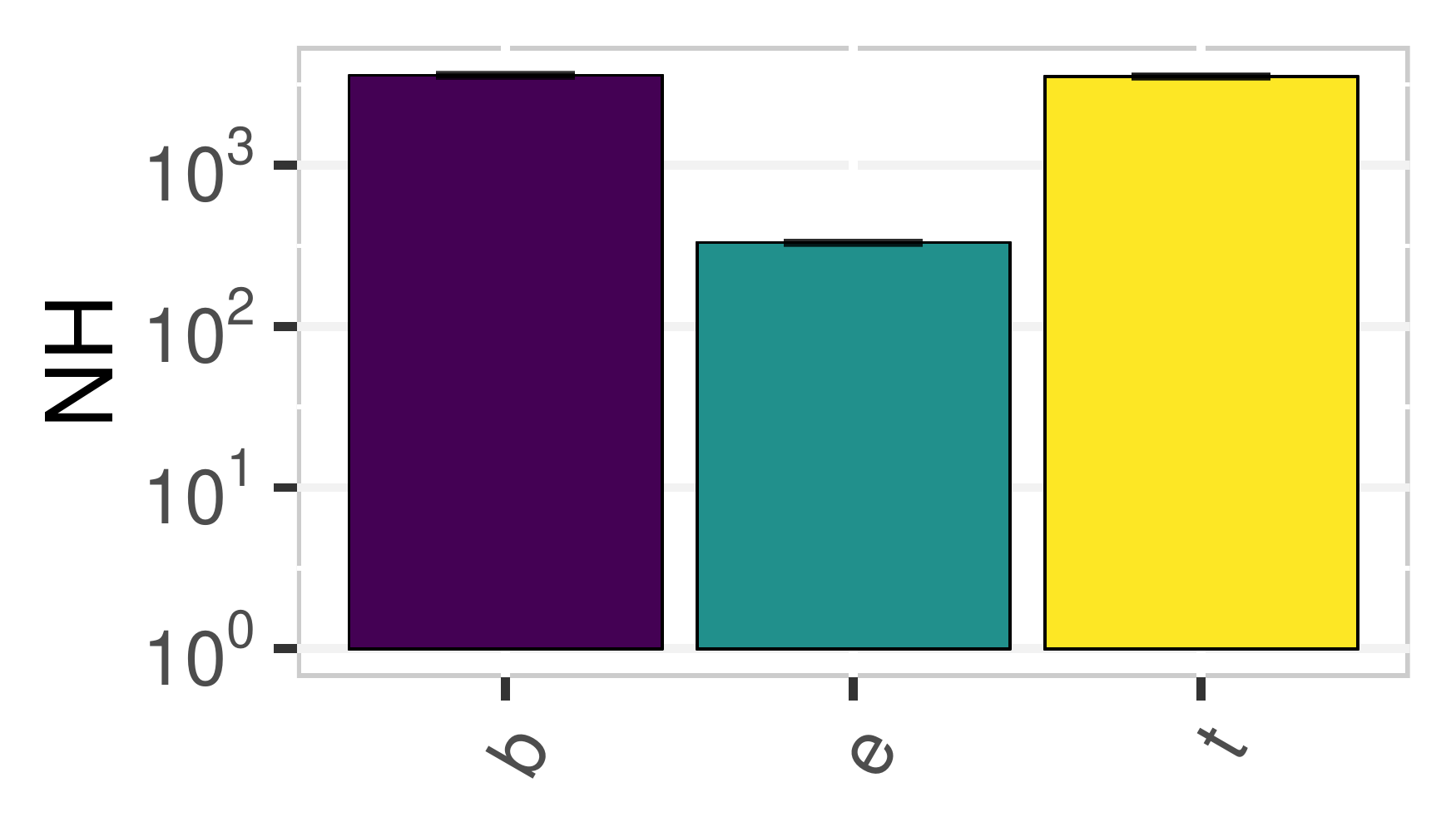}
  }
  \subfloat[NRKB]{
    \includegraphics[width=0.32\textwidth]{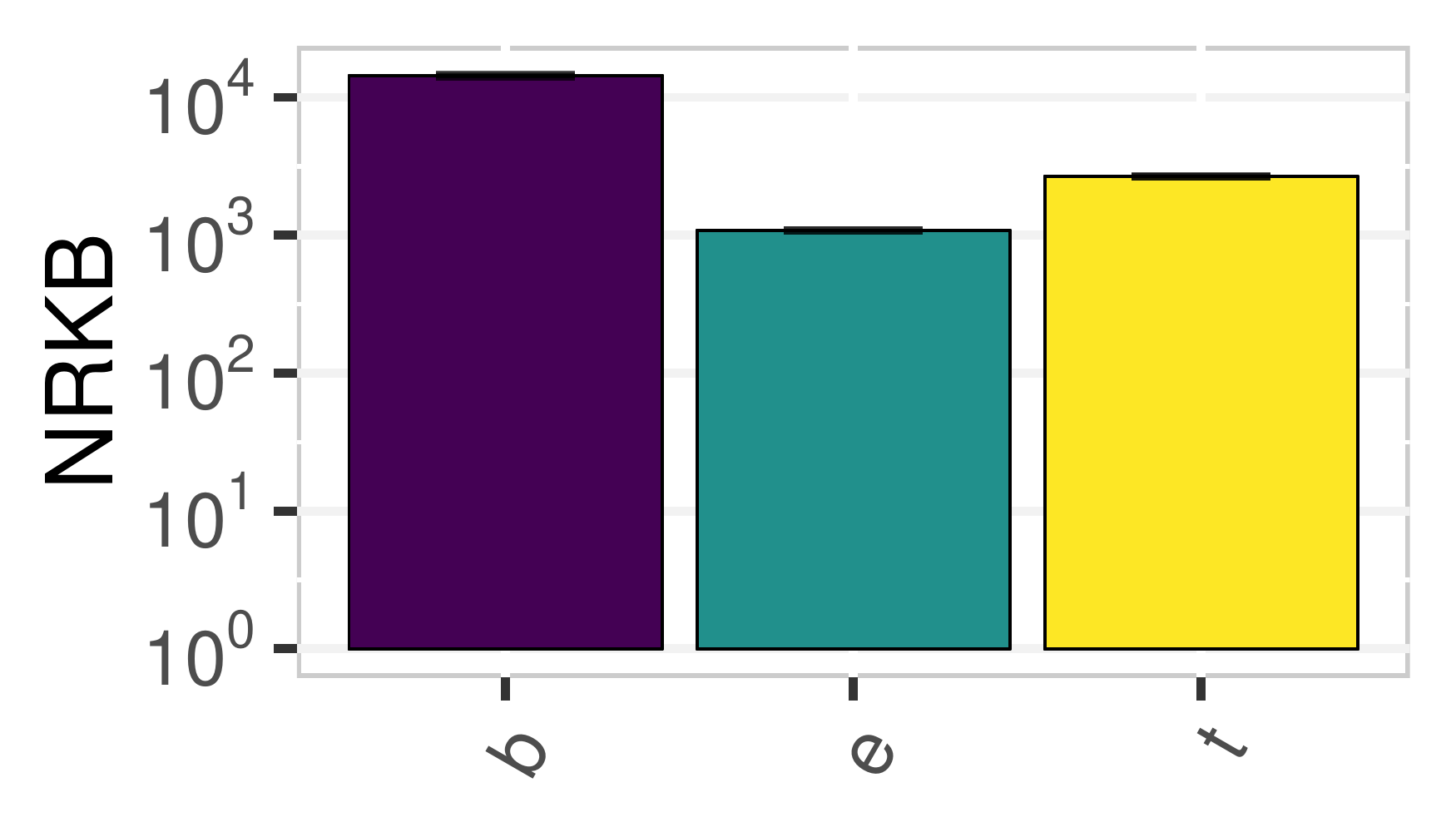}
  }
  \subfloat[SL/BL]{\label{fig:cpuUsage}
    \includegraphics[width=0.32\textwidth]{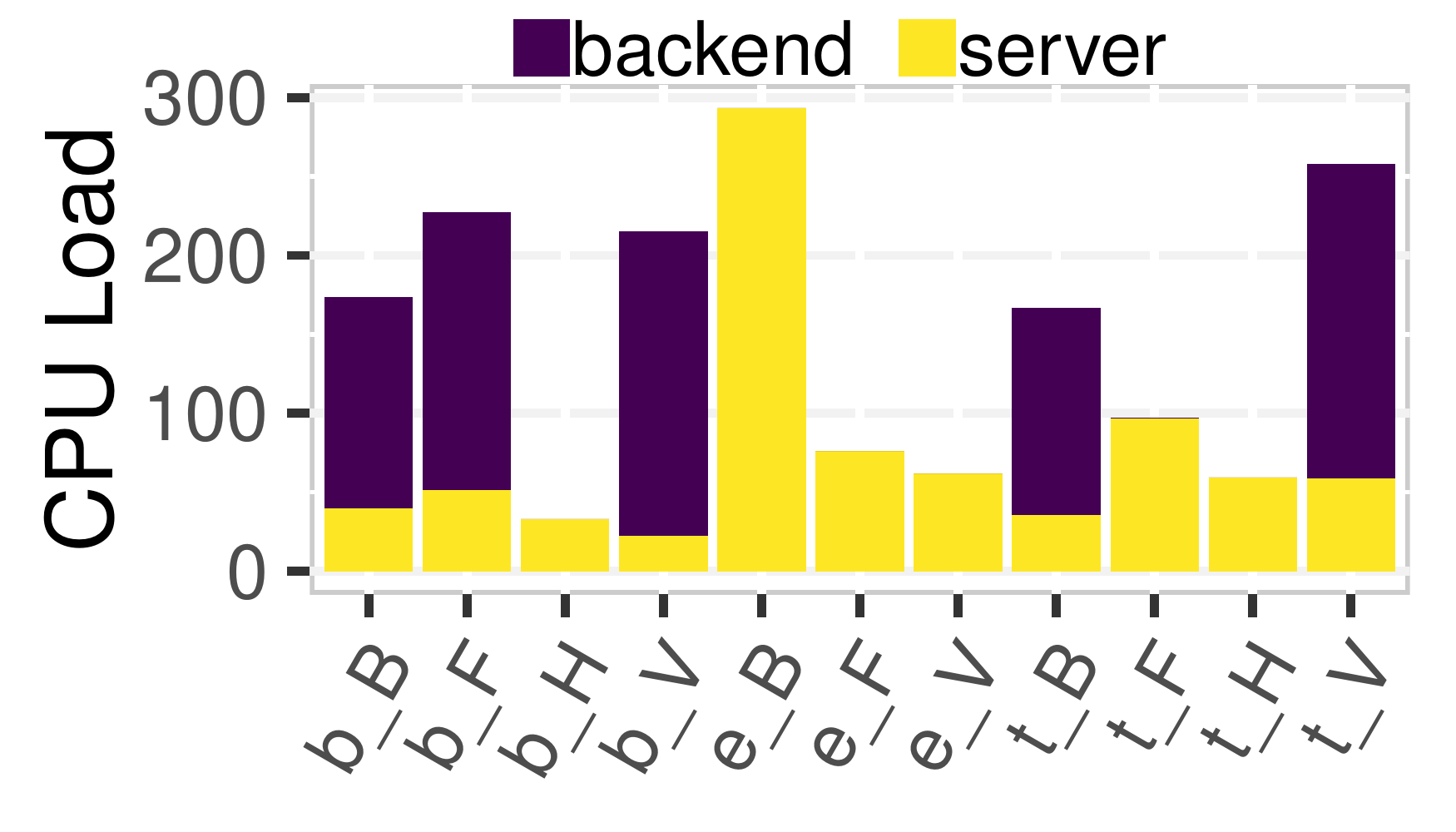}
  }
    \caption{NH, NRKB and CPU Load for Server and Backend (SL / BL)}
\vspace*{-2em}
\label{fig:tpfMetrics}
\end{figure*}

\textbf{Network Load. }
Figure~\ref{fig:tpfMetrics} shows the average number of requests (NH) and average amount of data transferred from the servers to the clients (NRKB) per interface as studied earlier~\cite{DBLP:conf/otm/HartigA16,DBLP:journals/ws/VerborghSHHVMHC16}. 
NH and NRKB are independent from the backend used; brTPF interface has the higher NH and NRKB, while the endpoint interface has the lowest. It is important to note that the data transfer from the server to the client is quite different from what has been shown earlier~\cite{DBLP:conf/otm/HartigA16}. In those experiments, brTPF is shown to decrease the network load between the server and the client with respect to both the number of HTTP requests and the amount of data transfer. However, we measured a higher network load with brTPF compared to TPF.

\begin{figure*}[!h]
\vspace*{-1.5em}
  \centering
  \subfloat[NRKB]{
    \includegraphics[width=\textwidth]{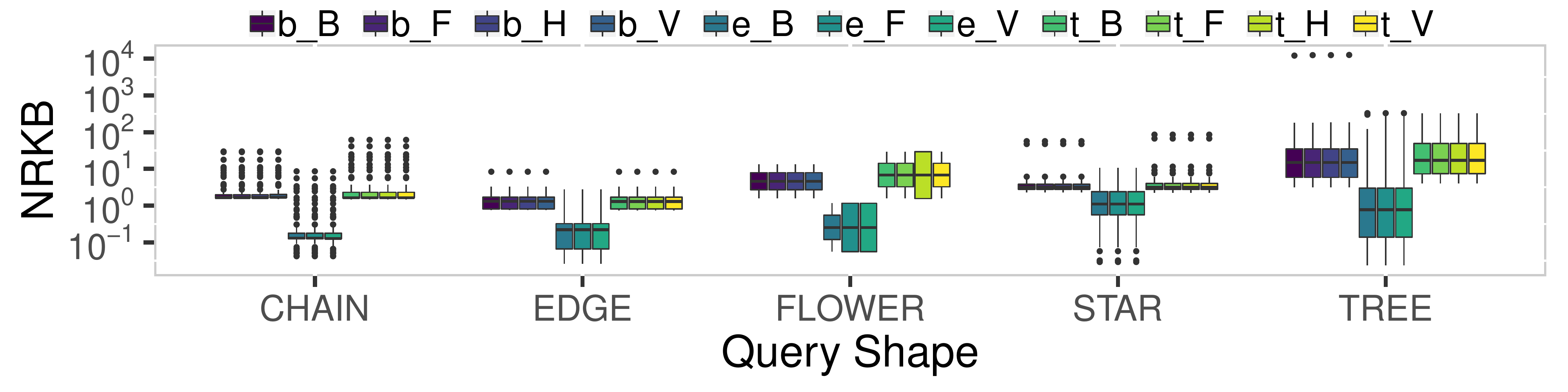}
  } \\
  \subfloat[NSKB]{
    \includegraphics[width=\textwidth]{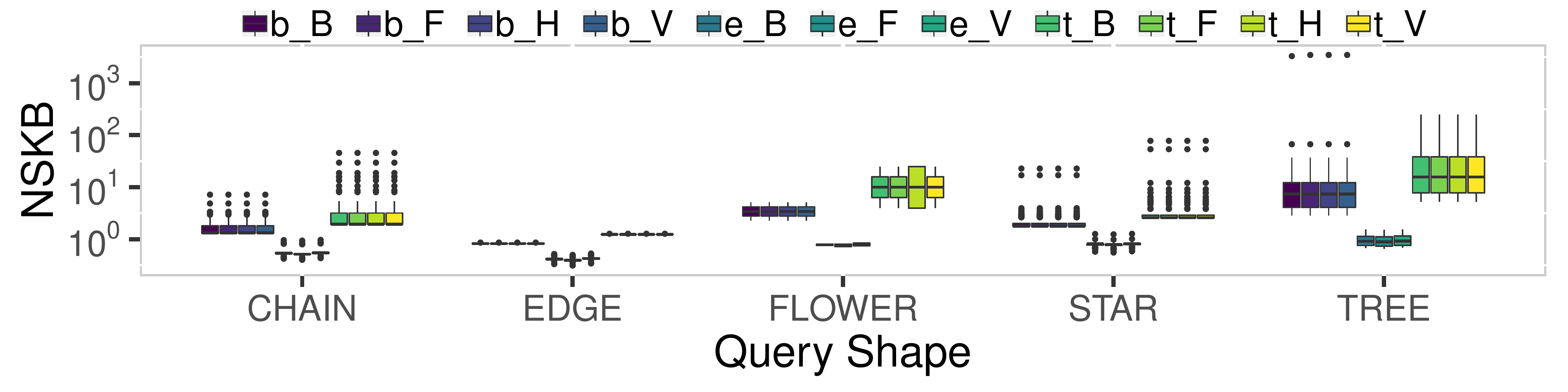}
  }\\
  \subfloat[NH]{
    \includegraphics[width=\textwidth]{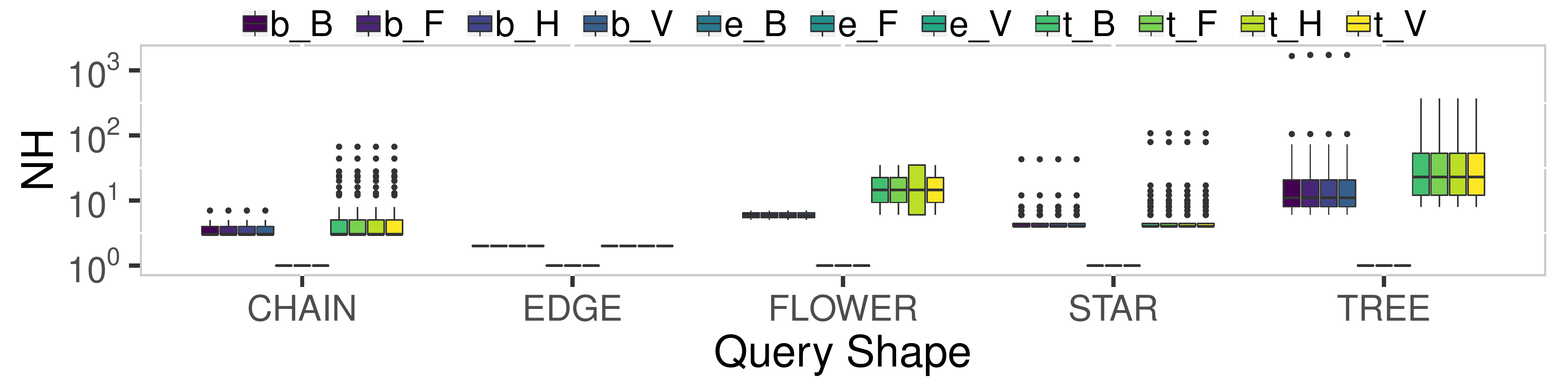}
  }
 	\vspace*{-0.5em}
    \caption{NRKB, NSKB, and NS per query load}
    \vspace*{-2em}
\label{fig:detailedMetrics}
\end{figure*}

Figure~\ref{fig:detailedMetrics} shows the number of transferred kBs from the server (NRKB) and from the client (NSKB) represented as a boxplot for each system and query load. Changing the backend for the interface does not introduce any changes for these metrics. The difference across the interfaces for NRKB is quite large for CHAIN-shaped queries and quite small for STAR-shaped queries. Processing STAR-shaped queries results in a higher data transfer for the endpoint interface, while it does not have a notable effect for TPF and brTPF interfaces. This  is the effect of the number of results in the case of the endpoint interface (Appendix~\ref{sec:appB} shows the number of results). The number of HTTP requests (NH) is constant and amounts to one for the endpoint interface as expected, while it is higher for the brTPF and TPF interfaces. Moreover, it increases with respect to the number of triple patterns included in the query for brTPF and TPF interfaces. In general, if we leave out the outliers (shown in the Figure~\ref{fig:detailedMetrics} as dots), an endpoint is the best performing interface and TPF is the worst performing interface for metrics related to the network load. This figure also illustrates that the outliers are the main reason why brTPF has a higher number of HTTP requests and a higher number of received kBs compared to TPF in Figure~\ref{fig:tpfMetrics} and the network load without considering outliers confirm the findings presented in~\cite{DBLP:conf/otm/HartigA16}. After performing evaluation on a finer granularity, we are able to see the effect of outliers, which would not be the case if we have only a single number to assess the interface.

\textbf{CPU Load. }
Figure~\ref{fig:cpuUsage} shows the CPU usage by the servers and their backends. While, overall, the endpoints use more CPU, using a Virtuoso endpoint uses less CPU than using a TPF server with Fuseki or HDT as backend. Moreover, if we consider the CPU used by the backend, then the Virtuoso endpoint uses less CPU than any of the TPF implementations. However, when Virtuoso is used as backend for TPF, it exhibits the highest CPU load for a TPF implementation. 
Additionally, the Virtuoso endpoint also exhibits a lower and better CPU usage than most brTPF combinations except the case where HDT is used as backend.
The CPU loads obtained from single-client experiments confirm the findings of earlier work~\cite{DBLP:journals/ws/VerborghSHHVMHC16, DBLP:conf/otm/HartigA16} since HDT file was used as a backend for both TPF and brTPF and this results in very low CPU load as we show in Figure~\ref{fig:cpuUsage}. However, CPU loads measured for TPF and brTPF interfaces with Virtuoso, Fuseki, and Blazegraph backends do not support the claim that TPF and brTPF interfaces would decrease the CPU load on the server side if SPARQL endpoints are used as backends. On the contrary, except Blazegraph, TPF and brTPF results in a higher total CPU load when using SPARQL endpoints as backends as shown in Figure~\ref{fig:cpuUsage}.

\subsection{Multiple Clients}
To stress the systems with multiple concurrent clients, we have executed experiments with $1$, $16$, $32$, and $64$ clients. We have processed Equal and Proportional query loads and report execution time (ET) that is the time elapsed since the beginning of processing the query loads until all the clients are done. In the figures with box-plots, we show the distribution of the metrics in the query load. The results illustrate the trade-offs between different interfaces and different backends. 

\begin{figure}[!h]
\vspace*{-1.5em}
  \centering
    \includegraphics[width=\textwidth]{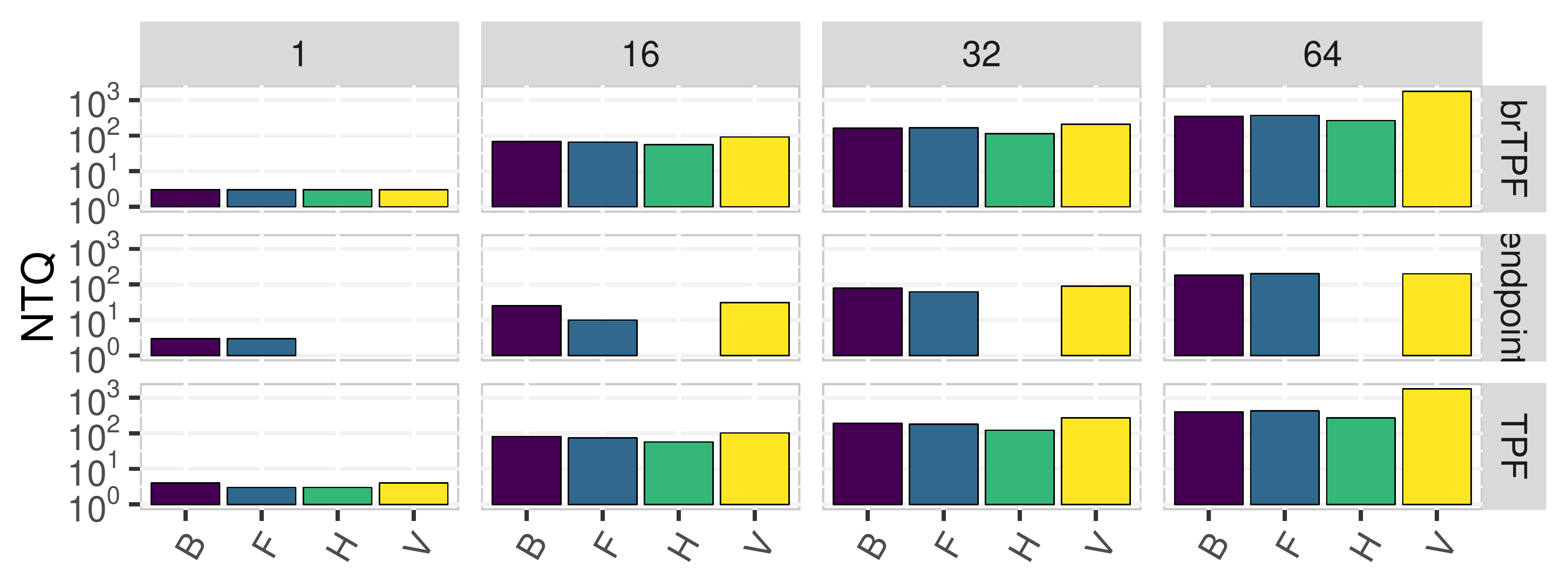}
	\vspace*{-1.5em}
    \caption{Total Number of Timed out Queries for 1, 16, 32, and 64 clients}
    \vspace*{-2em}
\label{fig:detailedNQT}
\end{figure}

\textbf{Performance. }
While TPF and brTPF are designed to reduce the server load, they also considerably increase the execution time of queries. brTPF and TPF result in higher numbers of query timeouts than the endpoints regardless of the chosen backend (see Figure~\ref{fig:detailedNQT}). Moreover, the number of timeouts is considerably higher for brTPF and TPF for 64 clients when Virtuoso is used as a backend. 

\begin{figure}[!h]
\vspace*{-2em}
  \centering
    \includegraphics[width=\textwidth]{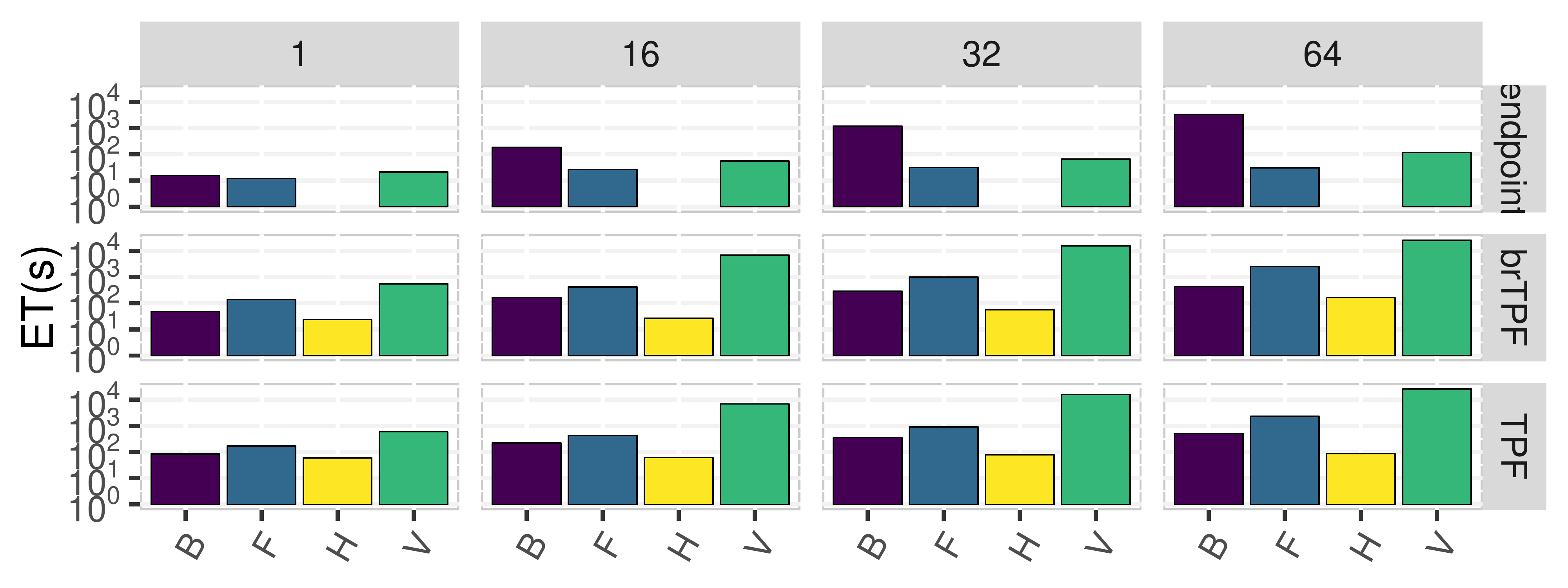}
    \vspace*{-1.5em}
    \caption{ET for 1, 16, 32, and 64 clients}
    \vspace*{-2em}
\label{fig:detailedETMC}
\end{figure}

Figure~\ref{fig:detailedETMC} shows the execution time as traditionally presented in existing studies~\cite{DBLP:conf/otm/HartigA16,DBLP:journals/ws/VerborghSHHVMHC16}. For the endpoint interface, Fuseki achieves the best performance while Blazegraph shows the worst performance. Interestingly, while Blazegraph is the one with the lowest performance for the endpoint interface, it has the highest performance when used as backend for TPF and brTPF. Moreover, we can see that there are no changes in the relative performance of the backends within the same interface. HDT and Blazegraph achieve better performance compared to Fuseki and Virtuoso when used as a backend for brTPF and TPF. 

\begin{figure}[!h]
\vspace*{-1.5em}
  \centering
    \includegraphics[width=\textwidth]{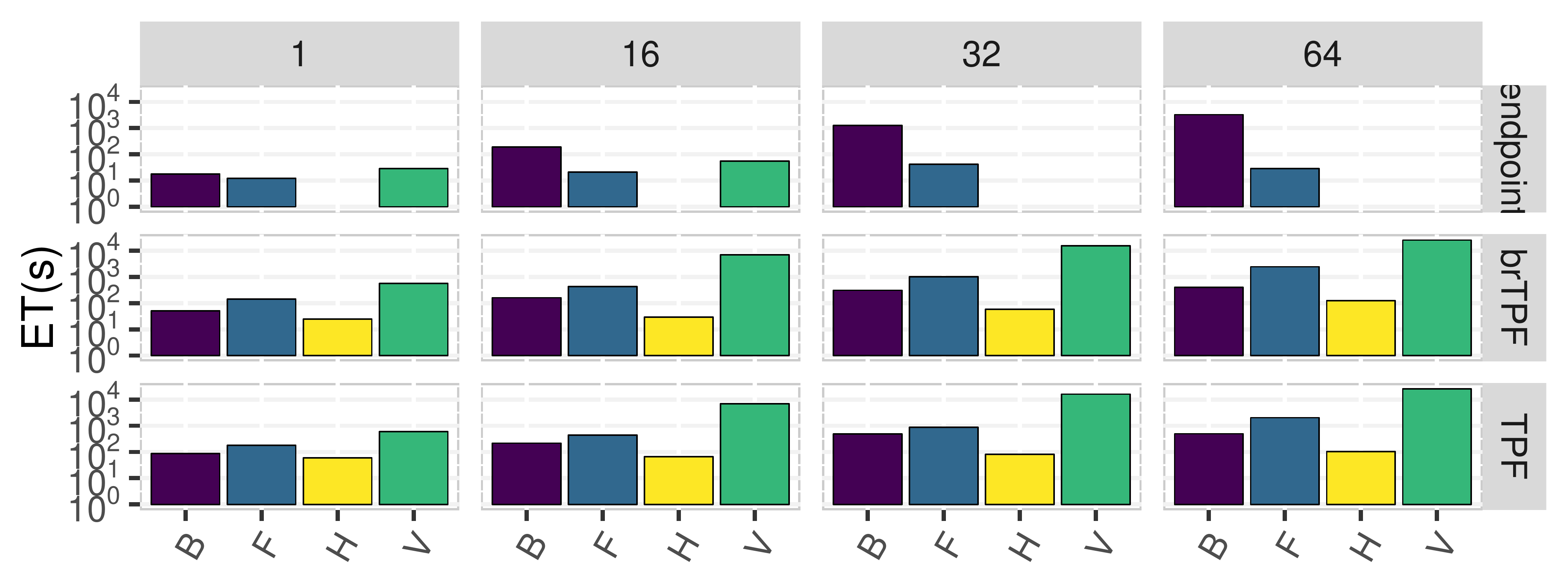}
    \vspace*{-2em}
    \caption{ET for 1, 16, 32, and 64 clients (8GB of main memory)}
    \vspace*{-2em}
\label{fig:detailedETMC8GB}
\end{figure}

Figure~\ref{fig:detailedETMC8GB} illustrates the execution time when each server is allocated $8$GB of main memory instead of $21$GB to assess whether the allocated memory to the server makes any difference. The execution times have a very similar trend compared to the execution times presented in Figure~\ref{fig:detailedETMC}. The only difference is that Virtuoso cannot handle $32$ and $64$ concurrent clients with $8$GB of main memory.

\begin{figure*}[htb]
  \vspace*{-1.5em}
  \centering
    \includegraphics[width=\textwidth]{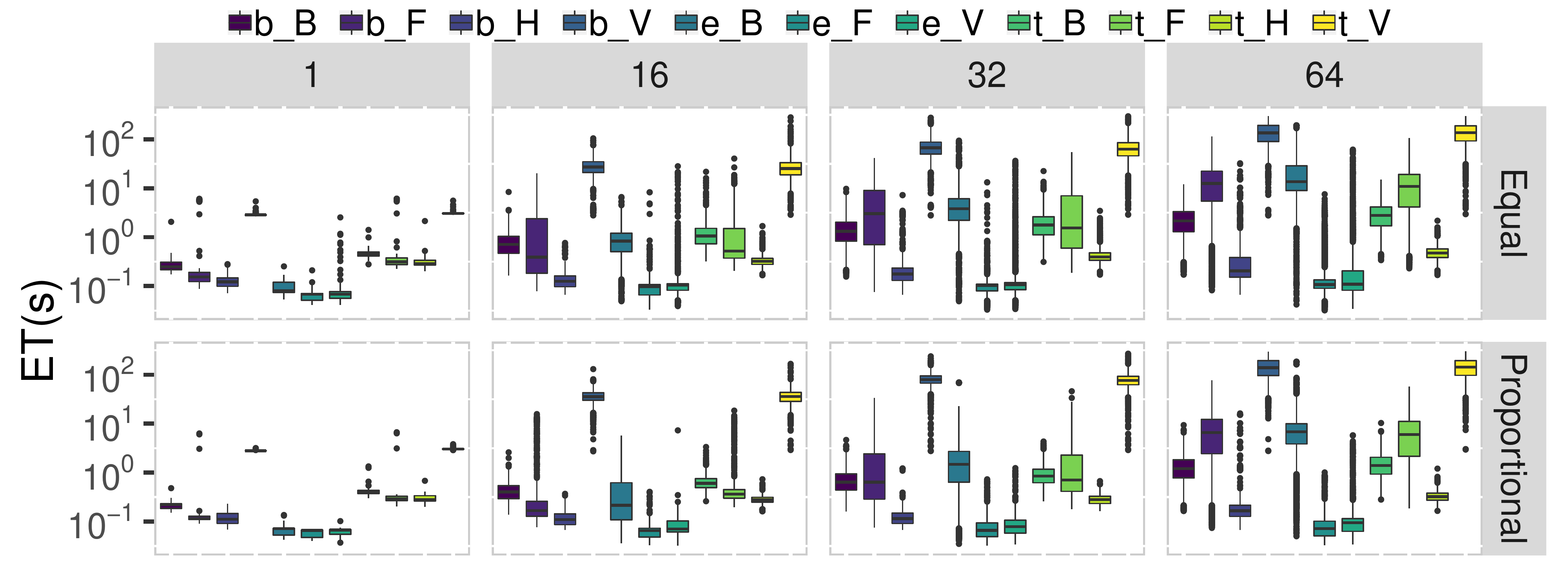}
    \vspace*{-1.5em}
    \caption{ET per approach and query load for 1, 16, 32, and 64 clients}
    \vspace*{-2em}
\label{fig:detailedET}
\end{figure*}

Figure~\ref{fig:detailedET} presents the execution time of the interface and backend combinations in a finer granularity for each query load. As shown in the figure, the endpoint produces the best performance when there is only a single client, which is in line with what we have seen with our single-client evaluation. However, when there are more than $16$ clients Fuseki, brTPF with HDT backend, and TPF with HDT backend perform better than the other combinations. Interestingly, TPF and brTPF with Virtuoso and Fuseki backends perform consistently worse than Virtuoso and Fuseki endpoints even with $64$ clients. In other words, using TPF and brTPF interfaces instead of an endpoint interface for these backends did not really have a positive effect on their query processing performance.

\textbf{Network Load. } Our experimental evaluation shows that there is no difference in the number of bytes the clients receive when the backend for an interface is changed (Appendix~\ref{sec:appC} shows the network load). Moreover, the difference between brTPF and TPF is very small. 

\begin{figure*}[htb]
  \centering
    \includegraphics[width=\textwidth]{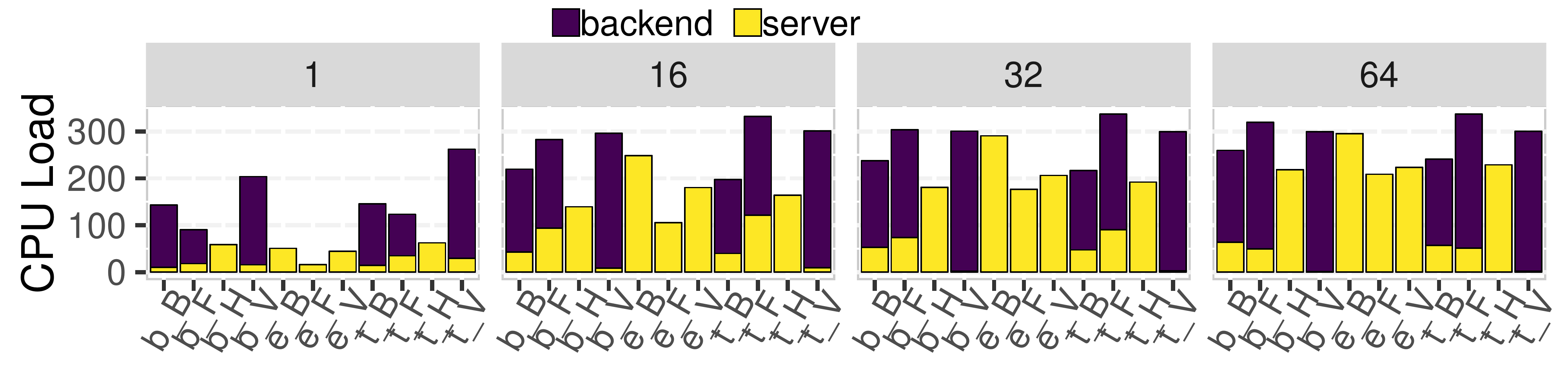}
    \vspace*{-1.5em}
    \caption{Average CPU usage for the servers and their backends for 1, 16, 32, and 64 clients}
    \vspace*{-2em}
\label{fig:detailedCPU}
\end{figure*}

\textbf{CPU Load. }
Figure~\ref{fig:detailedCPU} shows the average CPU loads of the server and the backends. All the systems have more CPU load as the number of clients increases, the endpoint interface is the one that is affected the most. The difference between the CPU loads of endpoints is quite significant between 1 client and 16 clients. This figure also demonstrates that the CPU load is lower when SPARQL endpoints are queried directly instead of being used as backend for TPF and brTPF. We can also conclude that brTPF and TPF do not provide a significant increase in availability of the SPARQL endpoints when they are used as backends except for Blazegraph. It is important to note that this has never been evaluated before. Moreover, we can safely conclude that HDT is the most suitable backend for brTPF and TPF.

\section{Conclusion}
\label{sec:conclusion}
In this paper, we presented an in-depth experimental evaluation of the state-of-the-art interfaces for querying linked data based on real query logs. We assessed the effect of query shapes on the performance of these interfaces. Moreover, we also examined the influence of backend selection on the performance.

The single-client evaluation results suggest that the shape of the query has a non-negligible effect on the performance of the interfaces. In addition, for complex query shapes like FLOWER and TREE, the endpoint interface provides the best performance in terms of execution time, network load, and CPU load. 

Our experiments clearly demonstrate that if the expected number of concurrent clients is not high, the endpoint interface appears to perform well. We also evaluated the network load for single-client and multi-client setups and our experimental evaluation shows that earlier findings only hold if we exclude very challenging queries such as the outliers in TREE query load. As future work, we want to analyze how such challenging queries affect the query processing performance. Moreover, the experiments demonstrate that the selection of backend for TPF and brTPF has a huge effect on the query processing performance and the server load even for single-client setup. Our experiments clearly indicate that the number of timed-out queries with brTPF and TPF is higher than that of the endpoint interface. We also show that query processing performances of Virtuoso and Fuseki endpoints are consistently better than the TPF and brTPF with Virtuoso and Fuseki backends. The same also applies to the server load. These two observations demonstrate that TPF and brTPF do not provide any improvement over the availability of these SPARQL endpoints. The only case that the CPU load and the performance is improved for a SPARQL endpoint when used as a backend with TPF and brTPF is Blazegraph. However, the current implementation of TPF and brTPF interfaces make assumptions regarding queries with OFFSET and LIMIT modifiers that Blazegraph does not fulfill. For this reason, one should check whether the system complies with the assumptions of TPF and brTPF before using the system as a backend for them. In addition, we are able to confirm the earlier findings that TPF and brTPF with HDT backend produce a comparable performance compared to the endpoint interface. However, one should be also aware that the Fuseki endpoint can also operate on HDT and has a lower execution time with nearly the same server load. 

\vspace*{2ex}
\noindent
\textbf{Acknowledgments. }
This research was partially funded by the Danish Council for Independent Research (DFF) under grant agreement no. DFF-4093-00301 and Aalborg University's Talent Management Programme.

\bibliographystyle{abbrv}
{
	\bibliography{references}
}

\appendix

\section{Effect of Including Queries with Consistent Answers}
\label{sec:appA}

\begin{figure}[htb]
  \centering
  \subfloat[All Queries]{
    \includegraphics[width=0.48\textwidth]{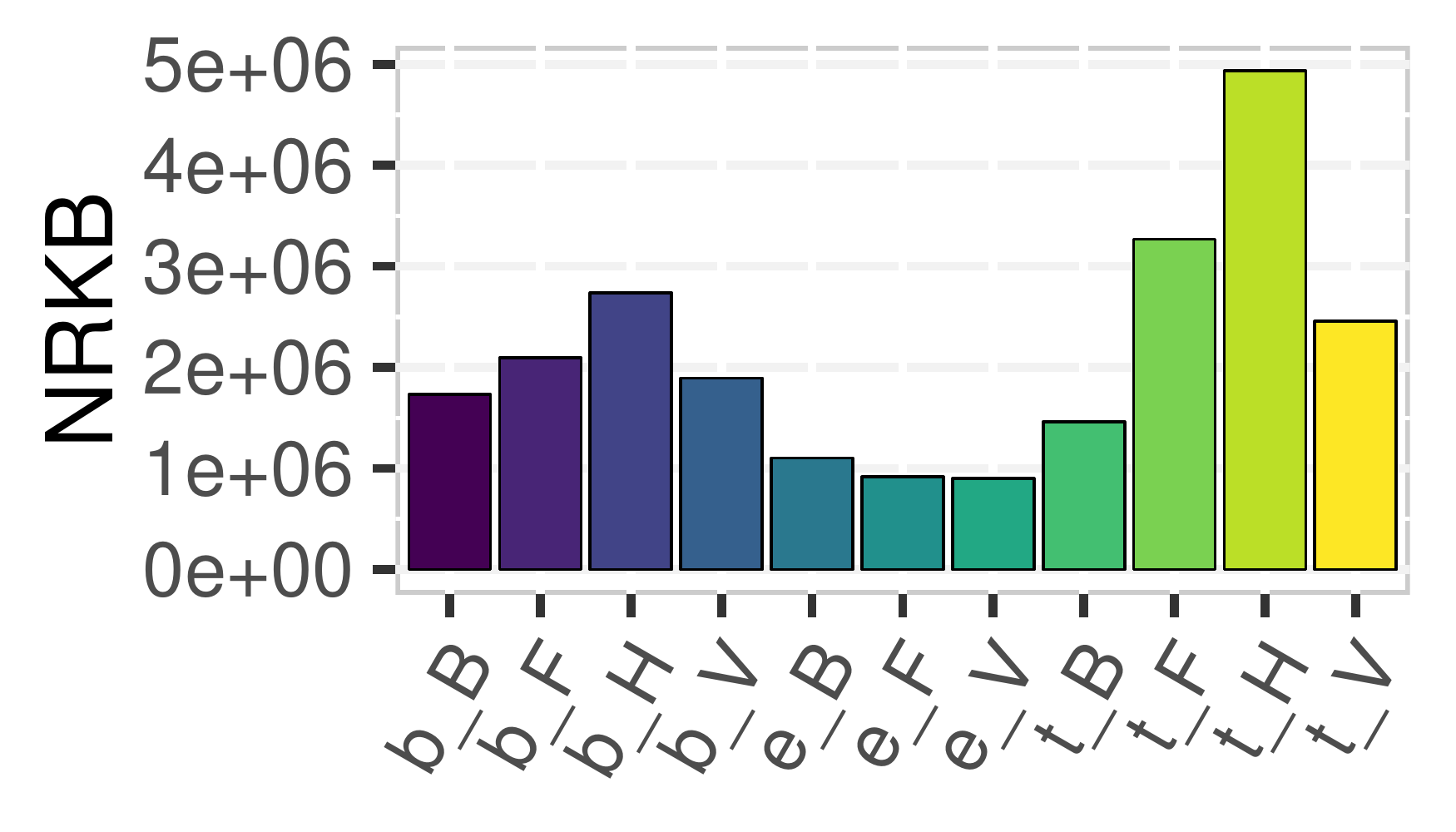}
  }
  \subfloat[Queries without Errors]{
    \includegraphics[width=0.48\textwidth]{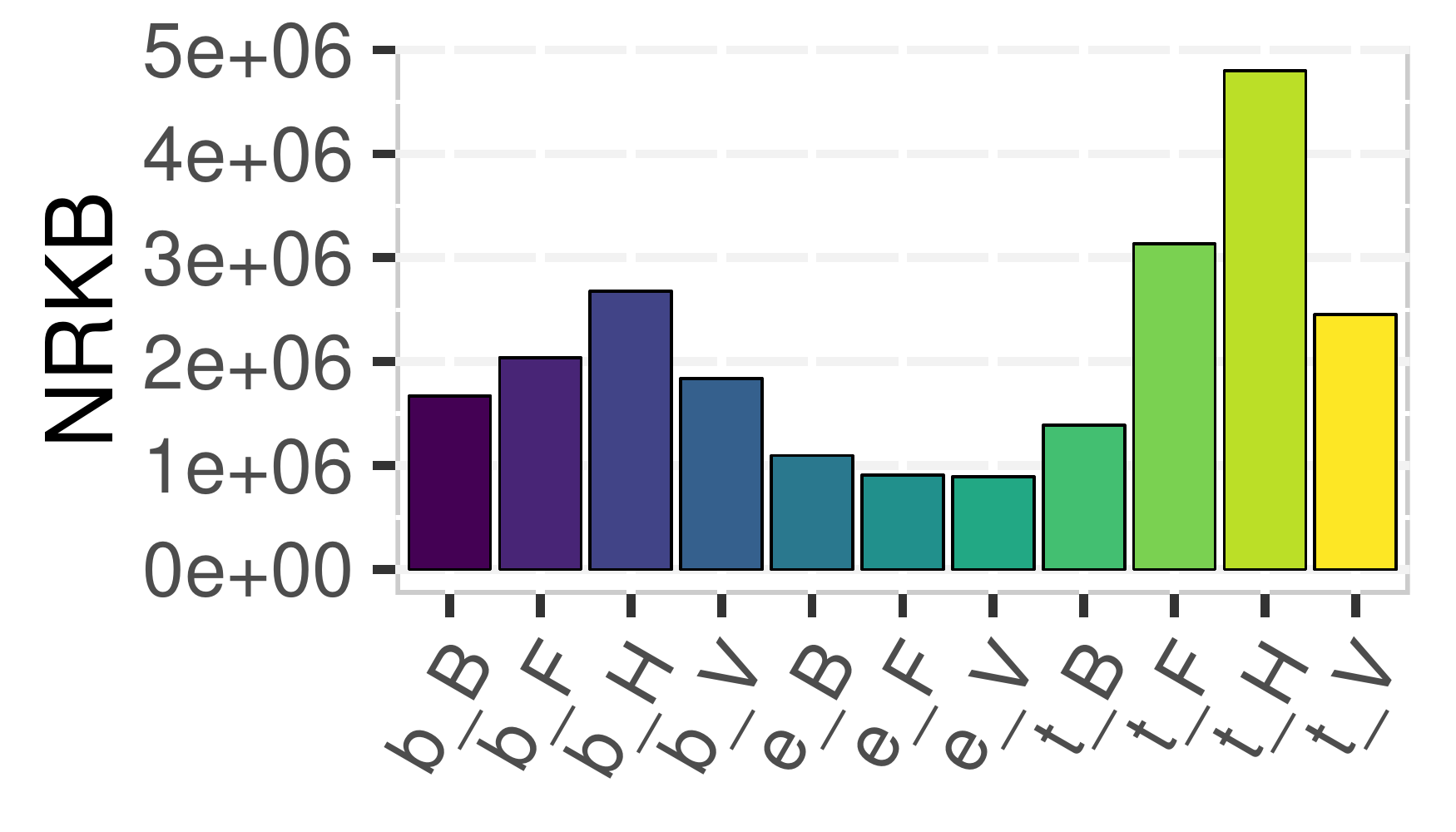}
  }\\
  \subfloat[Queries without Timeouts]{
    \includegraphics[width=0.48\textwidth]{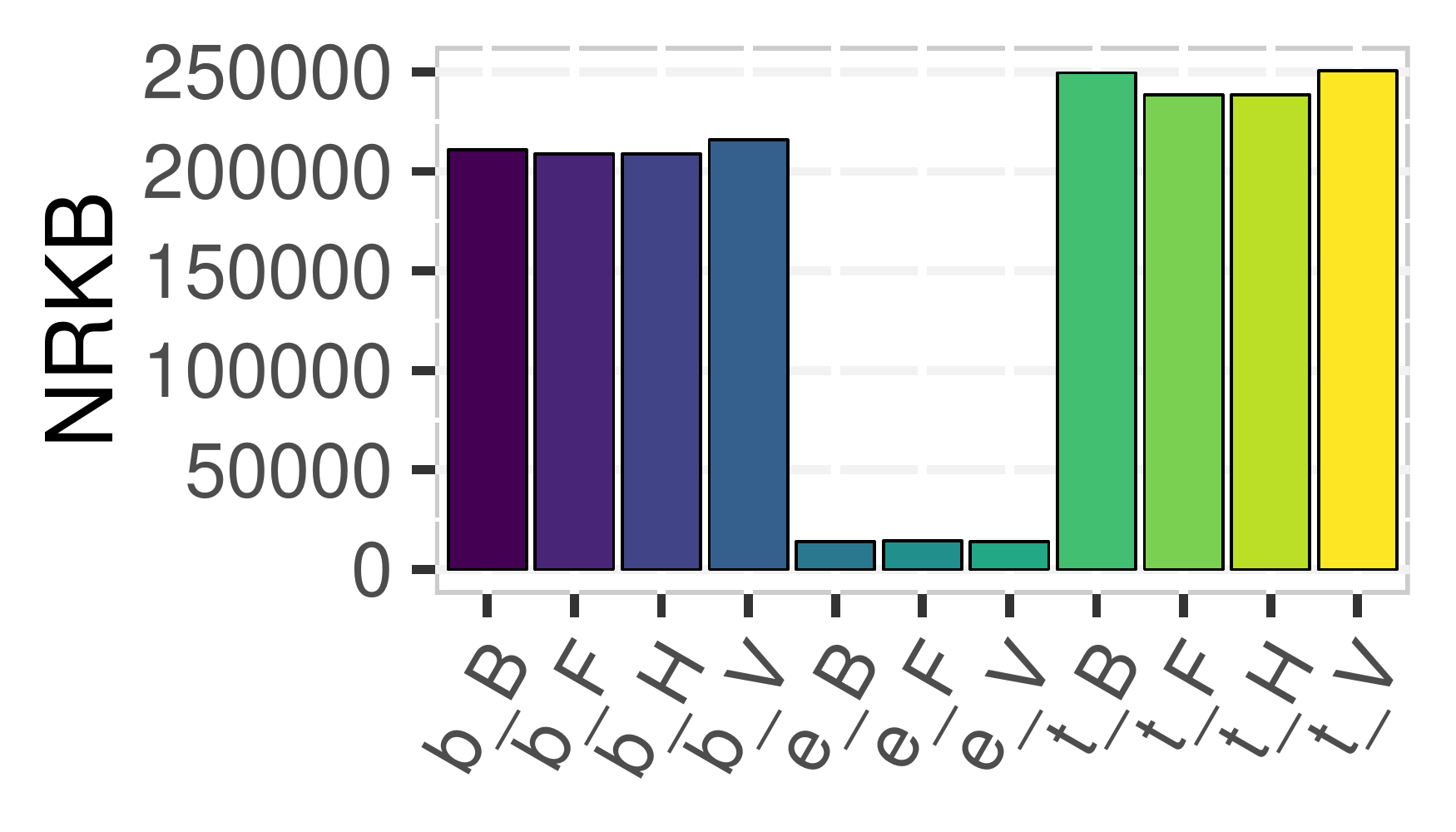}
  }
  \subfloat[Queries without Inconsistent Answers]{
    \includegraphics[width=0.48\textwidth]{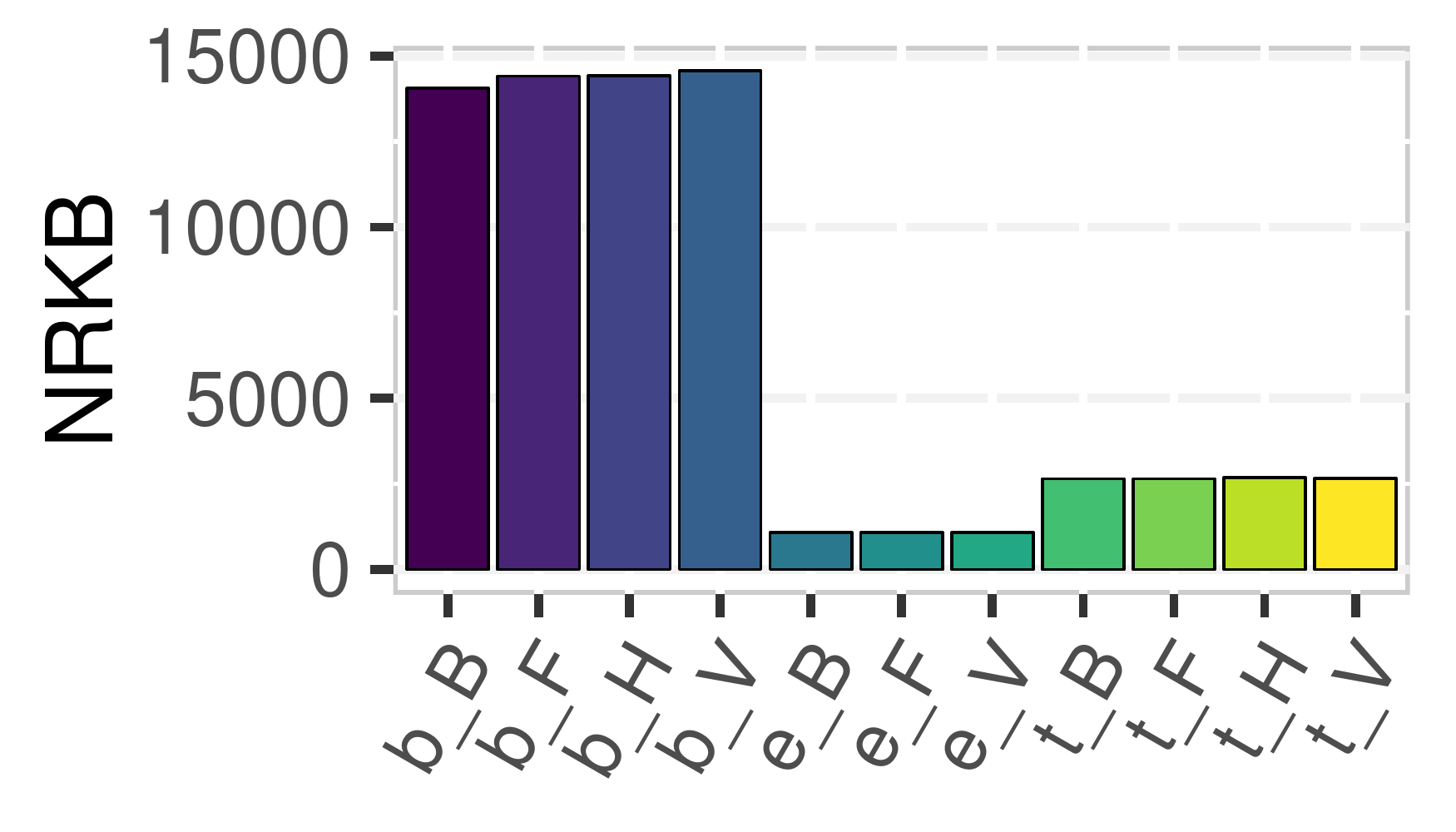}
  }
   \caption{Number of kB Received by the Client}
\end{figure}

\section{Number of Results}
\label{sec:appB}

\begin{figure}[htb]
  \centering
    \includegraphics[width=0.65\textwidth]{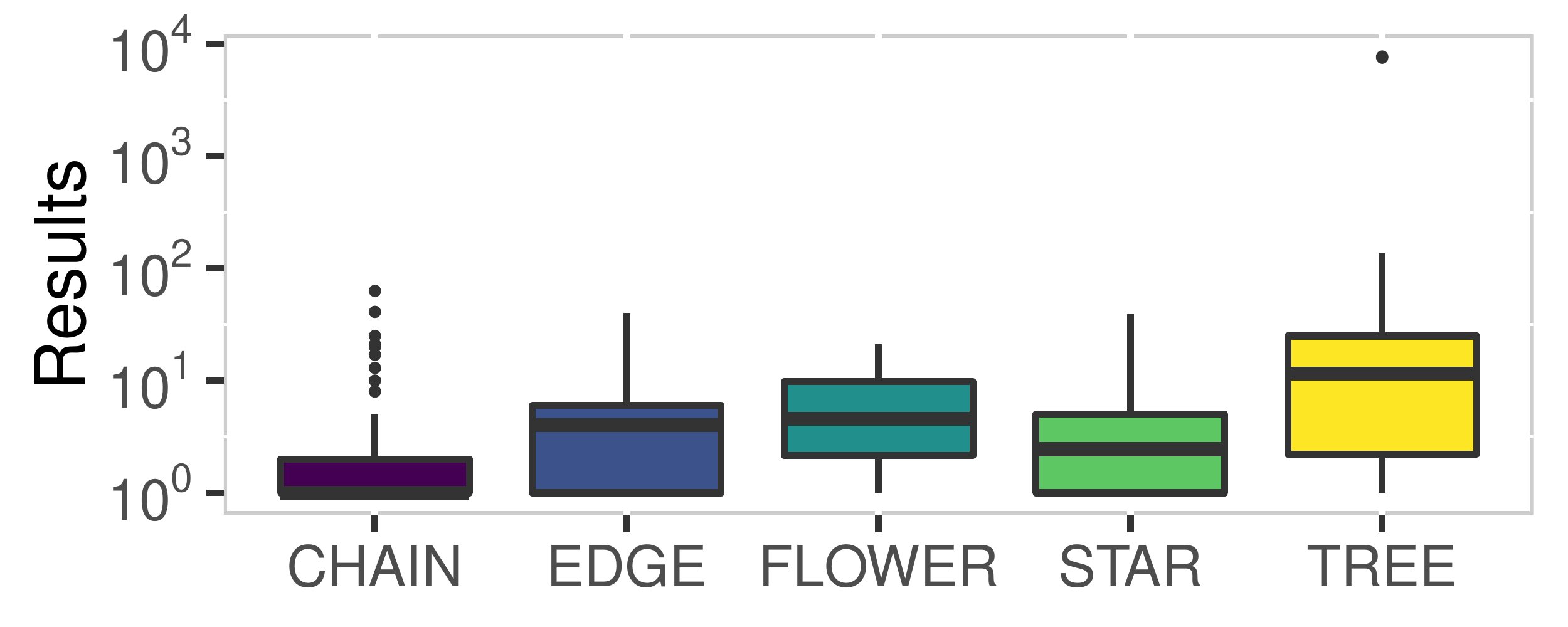}
   \caption{Number of Query Results per Shape}
\end{figure}

\section{Network Load}
\label{sec:appC}

\begin{figure}[htb]
  \centering
    \includegraphics[width=0.98\textwidth]{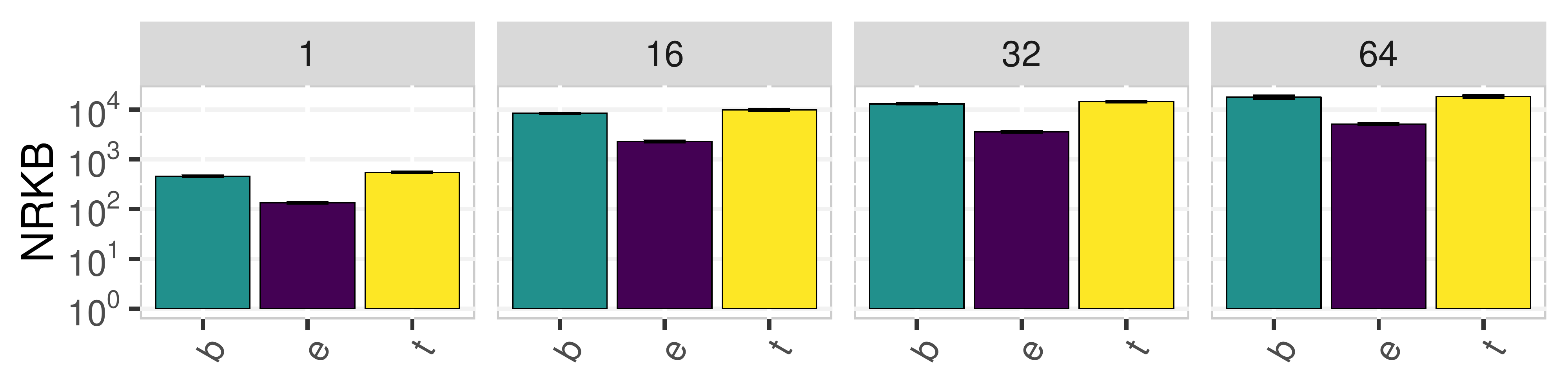}
   \caption{Average Number of kB Received by the Clients per Interface for 1, 16, 32, and 64 clients}
\end{figure}

\end{document}